\documentclass[a4paper,11pt]{article}
\usepackage[utf8]{inputenc}
\usepackage{hyperref}
\usepackage{amsmath}
\usepackage{mathtools}
\usepackage{amsfonts}
\usepackage{amssymb,amsthm}
\usepackage{graphicx}
\usepackage{epstopdf}
\usepackage{caption}
\usepackage{adjustbox}
\usepackage{subcaption}
\usepackage{afterpage}
\usepackage{lscape}
\usepackage{xcolor}
\usepackage{multirow}
\usepackage{cellspace}
\usepackage{breqn}
\newtheorem{theorem}{Theorem}
\newtheorem{prop}{Proposition}
\newtheorem{assump}{Assumption}
\newtheorem{corol}{Corollary}

\newtheorem{remark}{Remark}
\usepackage[font=small,labelfont=bf]{caption}
\usepackage{url}

\newcommand{\mb}[1]{\mathbf{#1}}
\newcommand{\bs}[1]{\boldsymbol{#1}}

\numberwithin{equation}{section}
\usepackage[toc]{appendix}

\usepackage[a4paper, total={6.5in, 9.5in}]{geometry}

\usepackage{authblk}
\usepackage[authoryear]{natbib}

\author{Giuseppe Buccheri\footnote{Department of Economics, University of Verona, giuseppe.buccheri@univr.it (corresponding author).}, Giacomo Bormetti\footnote{Department of Economics, University of Pavia.},  Fulvio Corsi\footnote{Department of Economics, University of Pisa.} and Fabrizio Lillo\footnote{Department of Mathematics, University of Bologna, and Class of Science, Scuola Normale Superiore, Pisa.}}

\renewcommand{\baselinestretch}{1.5}

\begin{document}

\newgeometry{left=2cm,right=2cm,top=1cm, bottom=2cm}

\title{Robust filtering and smoothing via perturbation methods\footnote{We acknowledge financial support from the Italian Ministry MUR under the PRIN project ``Dynamic models for a fast changing world:\ An observation-driven approach to time varying parameters" (grant agreement n.\ 20205J2WZ4).\ FL acknowledges partial support by the European Program scheme ``INFRAIA-01-2018-2019: Research and Innovation action", grant agreement \#871042 'SoBigData++: European Integrated Infrastructure for Social Mining and Big Data Analytics.}}

\date{}
\maketitle
\begin{center}
\vspace{-0.8cm}
December, 2024 \\
\end{center}
\vspace{0.2cm}

\renewcommand{\baselinestretch}{1.2}

\begin{abstract}

Using a perturbation technique, we derive a new approximate filtering and smoothing methodology generalizing along different directions several existing approaches to robust filtering based on the score and the Hessian matrix of the observation density.\ The main advantages of the methodology can be summarized as follows:\ (i) it relaxes the critical assumption of a Gaussian conditional distribution for the latent states underlying such approaches; (ii) can be applied to a general class of state-space models including location, scale and count data models; 
(iii) rationalizes the approximation to the likelihood function within the same perturbation approach, thus allowing for straightforward inference of the model parameters; (iv) enables the computation of confidence bands around the state estimates reflecting the combination of parameter and filtering uncertainty.\
We show through an extensive Monte Carlo study that the mean square loss with respect to exact simulation-based methods is small in a wide range of scenarios.\ We finally illustrate empirically the application of the methodology to the estimation of stochastic volatility and correlations in financial time-series.


\vspace{1cm}

\end{abstract}

\newpage

\newgeometry{left=2.5cm,right=2.5cm,top=2cm, bottom=2cm}

\renewcommand{\baselinestretch}{1.6}

\section{Introduction} 

Economic and financial time-series often exhibit significant dynamic properties that can be described using state-space models with time-varying latent states. In a linear and Gaussian state-space model, the Kalman filter provides a simple recursive solution to the problem of finding the optimal filtered and smoothed estimates of the state variables.\
However, many time-series models are nonlinear and/or non-Gaussian. Well-known examples include stochastic volatility and correlation models (\citealt{HarveySV}), stochastic duration and intensity models (\citealt{BAUWENS2004381}, \citealt{BauwensEtAl2006}), dynamic copula models (\citealt{HafnerHans2012}). In these cases, the filtered and smoothed estimates can only be computed using simulation-based methods, such as importance sampling, particle filters, Monte Carlo Markov-Chains; see, for example, \cite{ShephardPitt} and \cite{DoucetFreitasGordon}. 

The use of simulation-based methods naturally requires significant computational resources compared to the simple Kalman filter.\ Consequently, researchers often prefer adopting approximate simple solutions to the filtering problem rather than finding the optimal but computationally expensive solution.\ An important class of approximate filtering approaches that has emerged in both the statistical and engineering literature is the one using the score and the Hessian matrix of the observation density in order to robustify the Kalman filter.\ Early examples of such approaches date back to \cite{Masreliez}, \cite{Martin79}, \cite{West81} and plunge their roots in the work on robust estimation of \cite{Tukey60} and \cite{Huber64}.\ %
While extensive Monte Carlo and empirical evidence has been found showing that filtering methods based on the score and the Hessian matrix perform well in practice, they are derived under the crucial assumption of a Gaussian prior density for the latent states.\ As argued by \cite{SchickMitter94} and \cite{Harvey_2013}, such an assumption is clearly violated in nonlinear and/or non-Gaussian state-space models, and thus remains controversial and difficult to motivate.


In this paper, we generalize this class of approximate filtering methods along several directions.\ First, we derive the filtering recursions by relaxing the normality assumption for the prior density.\ Our results show that the use of the score and Hessian matrix as driving forces for the filter is robust under significantly weaker tail conditions for the prior distribution which are more likely to be satisfied in nonlinear and/or non-Gaussian state-space models.\ 
Second, while existing approaches derive the approximate recursions assuming a location model as a data generating process, we consider a much wider class of data generating processes including also scale, count data and other types of state-space models, thus extending significantly the range of possible applications.\ Third, our method approximates not only the conditional state estimates, but also the likelihood function.\ Therefore, we can obtain approximate estimates of the unknown model parameters within the same methodological framework.\  
Finally, we show how to compute approximate confidence bands around the filtered and smoothed estimates.\ 

Confidence bands are extremely important in empirical work because they quantify the uncertainty surrounding the time-varying parameters.\ However, their computation in nonlinear and non-Gaussian state-space models is generally problematic and relies on simulation-based methods; see e.g., \cite{shephard2005stochastic}.\ We derive in-sample and out-of-sample confidence bands reflecting the combination of two types of uncertainty:\ (i) filtering uncertainty, referring to the latent states uncertainty surviving upon conditioning on the sample data; (ii) parameter uncertainty, referring to the uncertainty affecting the maximum-likelihood estimates of the static parameters. We show that both kinds of uncertainty are relevant when computing confidence bands around the latent state dynamics.    
 
The methodology we propose has the same computational complexity of the Kalman filter, but guarantees a robust update of the state estimates in case of deviations from linearity and/or normality. The filtered and smoothed estimates of the latent states are indeed computed through a single forward-backward recursion, in a very similar fashion to the Kalman filter and smoother recursions.\ However, when the underlying model is nonlinear and/or non-Gaussian, the updating mechanisms for the conditional mean and variance are quite different compared to the standard Kalman filter. On the one hand, the score provides robust mean estimates by winsorizing the tails of the observation density. On the other hand, contrary to the Kalman filter, where the conditional variance is a deterministic function of time, in our approach it follows a stochastic recursion driven  by the Hessian of the observation density.  
This different structure is quite natural, considering that the independence of the conditional variance from the sample data is a specific property of linear and Gaussian models, which becomes sub-optimal in more general settings.\ In fact, the Hessian acts non-linearly on the sample data providing robust variance estimates when observations are unlikely to be generated  by a normal distribution.  

While the robustness provided by the score and Hessian is common to other approximate filtering methods, we show that it can be motivated under more general assumptions.\ The approximate filtering methodology is derived using a perturbation approach in the spirit of \cite{FlemingEwndell71}.\ Generally speaking, this class of approaches considers a particular case in which the solution to a given problem has a simple form, and obtains an approximate general solution by computing small deviations around this special case.\ Following this approach, we expand the observation density in power series around the particular solution corresponding to a degenerate prior distribution for the latent states.\ We then obtain the general approximate solution to the filtering  problem as a perturbation of this particular case.\ 
The error resulting from this local approximation can be formally characterized under a set of tail conditions for the prior density which are significantly weaker compared to the standard normality assumption made in the literature.\ 

We perform an extensive Monte Carlo analysis using different types of state-space models and comparing the proposed methodology with exact simulation-based methods and other approximate filtering techniques. Despite its computational simplicity, we find that the average mean square loss of our approximate methodology with respect to exact methods is smaller than 2\% in many of the scenarios considered in the analysis.\ Our empirical application, instead, aims to assess in a real data environment the improvement of the update and smoothing steps with respect to the prediction step. This is done by comparing the mean square distance between the (model based) (co)variance estimates obtained using daily returns, and the (model free) realized (co)variance measures constructed using intraday data. The analysis reveals that the use of contemporaneous and subsequent information significantly reduces the mean square distance. Furthermore, neglecting filtering or parameter uncertainty entails a number of exceedances larger than expected when constructing confidence bands around the filtered and smoothed estimates.


The rest of the paper is organized as follows. In Section \eqref{sec:filtSmooth}, we present the methodology and discuss its relation with other approximate filtering approaches based on the score and the Hessian matrix. In Section \eqref{sec:statProp}, we illustrate the perturbation approach allowing to derive the filtering and smoothing methodology.\ We also discuss in detail the regularity conditions allowing to characterize the approximation error and the construction of the confidence bands around the state estimates. Sections \eqref{sec:MonteCarlo}, \eqref{sec:empirics} present the simulation and empirical results, respectively. Finally, Section \eqref{sec:Conclusions} concludes. The proofs of the results in Section \eqref{sec:statProp} are reported in the appendix section.

\section{Methodology}

\subsection{Filtering and smoothing recursions}
\label{sec:filtSmooth}


Let $\mb{y}_t\in\mathbb{R}^p$ be a vector of observations and $\bs{\alpha}_t\in\mathbb{R}^m$ a vector of latent state variables. 
We consider a class of state-space model of the following form:\bigskip
\begin{align}
\mb{y}_t |\bs{\alpha}_t &\sim p(\mb{y}_t|\bs{\alpha}_t;\bs{\theta}) \label{eq:ssm_obs_gen}\\
\bs{\alpha}_{t+1} &= \mb{c}+ \mb{T}\bs{\alpha}_t+\bs{\eta}_{t+1}\medskip \label{eq:ssm_trans_gen}
\end{align} 

\noindent where $\{\bs{\eta}_t\}_{t=1}^n\in\mathbb{R}^m$ are i.i.d.\ innovations with zero mean and covariance $\mb{Q}\in\mathbb{R}^{m\times m}$, and $p(\mb{y}_t|\bs{\alpha}_t;\bs{\theta}) $ is a probability density with parameters $\bs{\theta}\in\mathbb{R}^q$.\ The system matrix $\mb{T}$ is assumed to be stable, meaning that all of its eigenvalues lie inside the unit complex circle.\ We clarify the assumptions on the observation density $p(\mb{y}_t|\bs{\alpha}_t;\bs{\theta})$ in Section \eqref{sec:statProp}. At this stage, it is sufficient to specify that $p(\mb{y}_t|\bs{\alpha}_t;\bs{\theta})$ is a (possibly) non-Gaussian density that depends on the state variables $\bs{\alpha}_t$ in a sufficiently smooth form.\ The specification in Equation \eqref{eq:ssm_obs_gen}, \eqref{eq:ssm_trans_gen} is quite general because it encompasses multivariate linear and nonlinear state-space models with Gaussian or non-Gaussian probability densities. For example, stochastic location, scale, copula, duration and count data models can be cast in this form. When $\bs{\alpha}_t$ is a location parameter and both $p(\mb{y}_t|\bs{\alpha}_t;\bs{\theta})$ and $p(\bs{\alpha}_{t+1}|\bs{\alpha}_t)$ are Gaussian, the above system of equations reduces to a linear and Gaussian representation whose optimal filtered and smoothed estimates can be recovered using the Kalman filter; see e.g., \cite{anderson1979} and \cite{Harvey}.

Let us denote by $\mb{Y}_t=\{\mb{y}_1,\dots,\mb{y}_t\}$ the information set at time $t$. The first two conditional moments of the latent states are defined as:\bigskip
\begin{align*}
\mb{a}_{t+1}=\mathbb{E}[\bs{\alpha}_{t+1}|\mb{Y}_t],\quad & \mb{P}_{t+1}=\mathbb{V}[\bs{\alpha}_{t+1}|\mb{Y}_t]\\
\mb{a}_{t|t}=\mathbb{E} [\bs{\alpha}_t|\mb{Y}_t],\quad & \mb{P}_{t|t}=\mathbb{V} [\bs{\alpha}_t|\mb{Y}_t]\\
\mb{a}_{t|n}=\mathbb{E}[\bs{\alpha}_{t}|\mb{Y}_n],\quad & \mb{P}_{t|n}=\mathbb{V}[\bs{\alpha}_{t}|\mb{Y}_n]\vspace{1cm}
\end{align*}  

\noindent with $t=1,\dots,n$. In the filtering literature, such conditional moments are known as predictive filtered estimates ($\mb{a}_{t+1}$, $\mb{P}_{t+1}$ ), update filtered estimates ($\mb{a}_{t|t}$, $\mb{P}_{t|t}$), and smoothed estimates ($\mb{a}_{t|n}$, $\mb{P}_{t|n}$). Let us also define the score and the Hessian matrix of the observation density as follows:\bigskip
\begin{equation}
\bs{\nabla}_t^{(i)} = \frac{\partial\log p(\mb{y}_t|\bs{\alpha}_t;\bs{\theta})}{\partial\bs{\alpha}_t^{(i)}}',\qquad  \bs{\mathcal{H}}_t^{(ij)} = \frac{\partial^2 \log p(\mb{y}_t|\bs{\alpha}_t;\bs{\theta})}{\partial \bs{\alpha}_{t}^{(i)}\partial\bs{\alpha}_{t}^{(j)}} \medskip
\label{eq_scHess}
\end{equation}
\noindent for $i,j=1,\dots,m$. 

The filtering and smoothing methodology we propose is based on the following forward-backward recursions providing approximate estimates of the first and second conditional moments of the latent states:
\begin{align}
\mb{a}_{t|t} = \mb{a}_t + \mb{P}_t\bs{\nabla}_t(\mb{a}_t),\qquad \qquad & \mb{P}_{t|t} = \mb{P}_t + \mb{P}_t\bs{\mathcal{H}}_t(\mb{a}_t)\mb{P}_t \label{eq:upd} \\
\mb{a}_{t+1}  = \mb{c}+ \mb{T}\mb{a}_t + \mb{T}\mb{P}_t\bs{\nabla}_t(\mb{a}_t), \qquad \qquad & \mb{P}_{t+1} = \mb{T}\mb{P}_t[\mb{T}+\mb{T}\mb{P}_t\bs{\mathcal{H}}_t(\mb{a}_t)]'+\mb{Q} \label{eq:pred}
\end{align}
for $t=1,\dots n$ and for some initial values $\mb{a}_1$, $\mb{P}_1$,
\begin{align}
\mb{r}_{t-1} = \bs{\nabla}_t(\mb{a}_t) + [\mb{I}+\mb{P}_t\bs{\mathcal{H}}_t(\mb{a}_t)]'\mb{T}'\mb{r}_{t}, \quad  & \mb{N}_{t-1} = -\bs{\mathcal{H}}_t(\mb{a}_t) + [\mb{I}+\mb{P}_t\bs{\mathcal{H}}_t(\mb{a}_t)]'\mb{T}'\mb{N}_{t}\mb{T}[\mb{I}+\mb{P}_t\bs{\mathcal{H}}_t(\mb{a}_t)]\nonumber\\
\mb{a}_{t|n}  =\mb{a}_t + \mb{P}_t\mb{r}_{t-1}, \quad & \mb{P}_{t|n} = \mb{P}_t-\mb{P}_t\mb{N}_{t-1}\mb{P}_t \label{eq:smooth}
\end{align}
for $t=n,\dots,1$ and for $\mb{r}_n=\mb{0}$, $\mb{N}_n=\mb{0}$. The notations $\bs{\nabla}_t(\mb{a}_t)$, $\bs{\mathcal{H}}_t(\mb{a}_t)$ indicate that $\bs{\nabla}_t$ and $\bs{\mathcal{H}}_t$, as defined in Equation \eqref{eq_scHess}, are computed for $\bs{\alpha}_t=\mb{a}_t$.\ 

The above filtering and smoothing recursions are derived in Section \eqref{sec:statProp} using a perturbation approach, which also allows us to recover an approximate likelihood function that can be used to infer the model parameters.\ To interpret such recursions, observe that, if the state-space model in Equation \eqref{eq:ssm_obs_gen}, \eqref{eq:ssm_trans_gen} is linear and Gaussian, i.e., $p(\mb{y}_t|\bs{\alpha}_t;\bs{\theta})=N(\mb{Z}\bs{\alpha}_t, \mb{R})$, where $\mb{Z}\in\mathbb{R}^{p\times m}$ and $\mb{R}\in\mathbb{R}^{p\times p}$ is a covariance matrix, then the score and Hessian matrix are given by
\begin{equation*}
\bs{\nabla}_t(\mb{a}_t)=\mb{Z}'\mb{R}^{-1}(\mb{y}_t-\mb{Z}\mb{a}_t),\quad \bs{\mathcal{H}}_{t}(\mb{a}_t)=-\mb{Z}'\mb{R}^{-1}\mb{Z} .
\end{equation*}
This implies that the conditional mean depends linearly on the prediction error $\mb{y}_t-\mb{Z}\mb{a}_t$, and that the conditional variance is independent from the sample data.\ Therefore, in a linear and Gaussian state-space model, the proposed algorithm behaves as the Kalman filter and smoother.\  However, the two filters do not coincide when the parameters are estimated using the exact likelihood function provided by the Kalman filter.\ They do coincide when, in our method, we estimate the parameters using the approximate Gaussian log-likelihood  $\sum_{t=1}^n \log p(y_t|\bs{\alpha}_t=\mb{a}_t;\bs{\theta})$, which arises from the same perturbation technique used to derive the filtering recursions; see Proposition \eqref{prop:pyt} below.\ Therefore, in a linear and Gaussian setting, our algorithm reduces to the classical Kalman filter and smoother recursions provided that \textit{both} the latent states and the static parameters are estimated using our perturbation technique; see also Remark \eqref{remark:KF} in Section \eqref{sub:approxResults}.\ 

\begin{figure}[htb]
    \centering 
\begin{subfigure}{0.5\textwidth}
  \includegraphics[width=\linewidth]{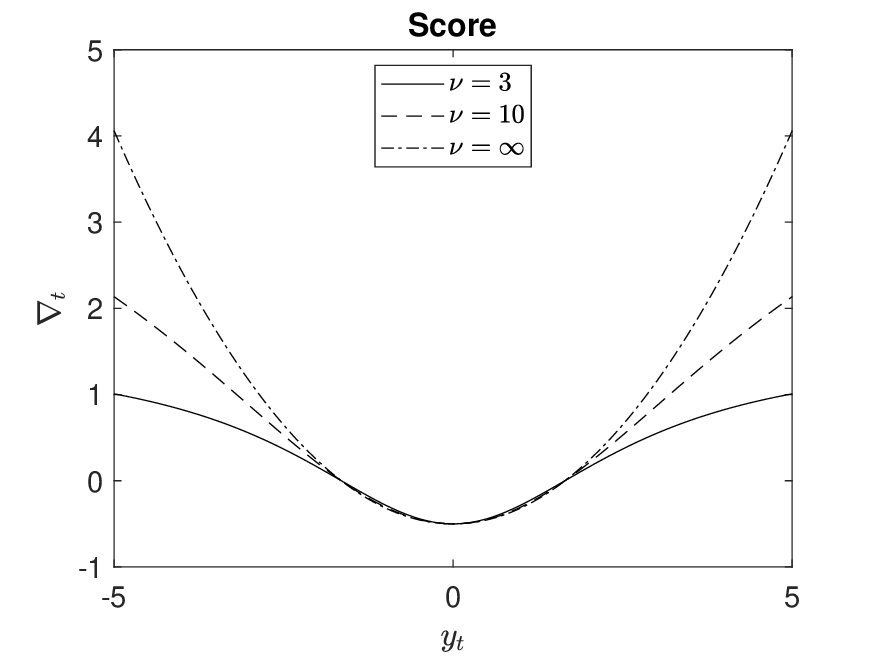}
\end{subfigure}\hfil 
\begin{subfigure}{0.5\textwidth}
  \includegraphics[width=\linewidth]{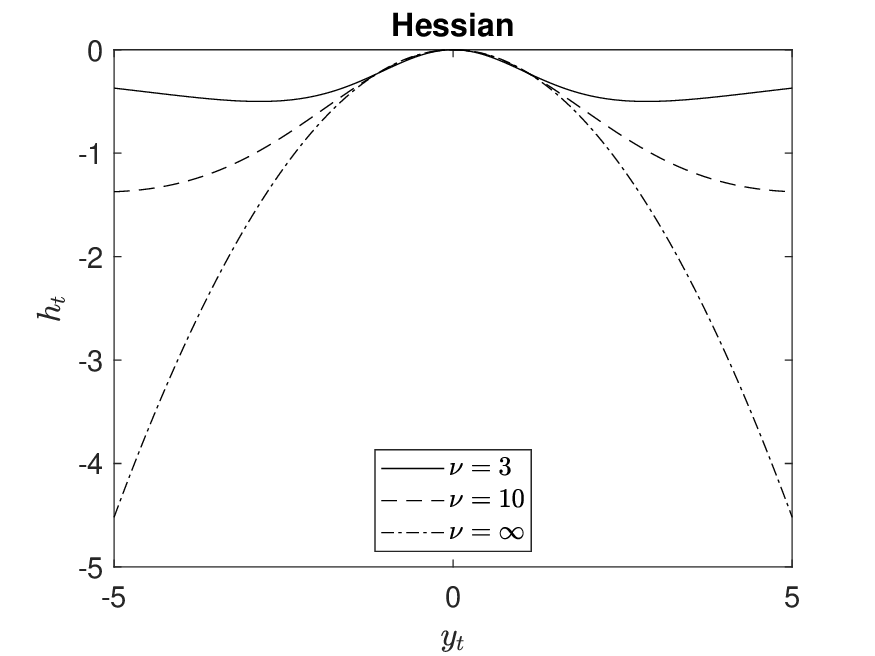}
\end{subfigure}
\caption{We plot, as a function of $y_t$, the score (left) and the Hessian (right) of the Student-$t$ scale model in Equation \eqref{eq:scaleStudUniMeas}, \eqref{eq:scaleStudUniTrans}, for $a_t=1$ and $\nu=3,10,\infty$. }
\label{fig:hessianScore}
\end{figure}

In a more general setting where $p(\mb{y}_t|\bs{\alpha}_t;\bs{\theta})$ is non-Gaussian, the conditional mean depends non-linearly on $\mb{y}_t$ and the conditional variance depends on the sample data. For example, let us consider a univariate scale model with Student-$t$ distribution:\medskip
\begin{align}
y_t &= e^{\frac{\alpha_t}{2}}\epsilon_t\label{eq:scaleStudUniMeas}\\
\alpha_{t+1} &= c + \phi\alpha_t +\eta_{t+1}\label{eq:scaleStudUniTrans}
\end{align}
where $\{\epsilon_t\}_{t=1}^n$ is an i.i.d.\ sequence of Student-$t$ innovations with $\nu$ degrees of freedom. The score is given by:
\begin{equation}
\nabla_t(a_t)= \frac{1}{2}\left[\frac{(\nu+1)y_t^2}{\nu e^{a_t}+y_t^2}-1\right],\medskip
\end{equation}
whereas the Hessian reduces to the second partial derivative of $\log p(y_t|\alpha_t;\nu)$, namely:\medskip
\begin{equation}
h_t(a_t)= -\frac{1}{2}\frac{\nu(\nu+1)e^{a_t}y_t^2}{(\nu e^{a_t}+y_t^2)^2}.\medskip
\end{equation}
The dependence of $\nabla_t$ and $h_t$ on the sample data is graphically represented in Figure \eqref{fig:hessianScore}, which plots both quantities as a function of $y_t$, for $a_t=1$ and $\nu=3,10,\infty$. When $\nu$ is low, e.g., $\nu=3$, the score and the Hessian downweight very large (in absolute value) observations.\ This has two main implications.\ First, the mean estimates, which are updated based on the score, are not very sensitive to such large observations. This is consistent with the fact that a large $|y_t|$ could be imputable to the fat-tail behavior of $p(y_t|\alpha_t;\theta)$.\ Second, since the variance estimates depend on $h_t$, which also downweights the extremely large observations, the latter are not useful to reduce our uncertainty on the state estimates. This is still consistent with the fact that, under non-normality, an extremely large observation is not informative about the state dynamics. Therefore, in the variance recursion, the Hessian has the same kind of winsorizing effect of the score in the mean recursion. 


\begin{figure}[htb]
    \centering 
\begin{subfigure}{0.5\textwidth}
  \includegraphics[width=\linewidth]{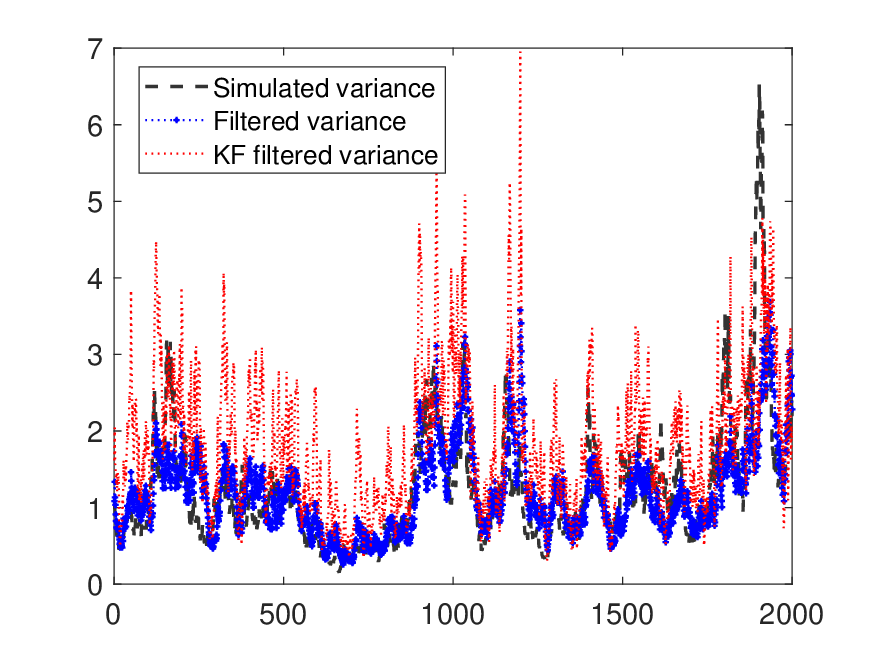}
\end{subfigure}\hfil 
\begin{subfigure}{0.5\textwidth}
  \includegraphics[width=\linewidth]{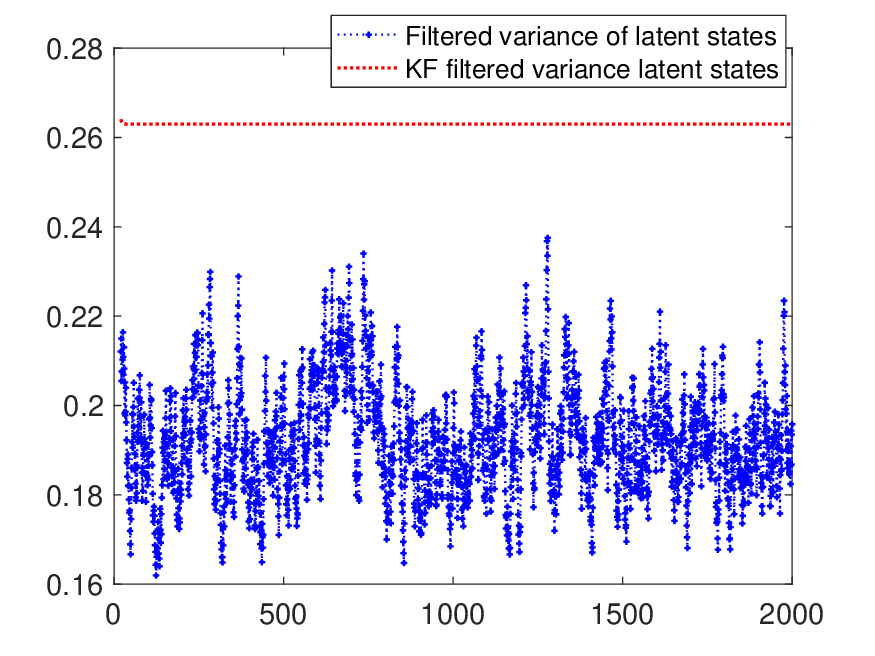}
\end{subfigure}
\caption{Left: simulated scale factor $\{e^{\alpha_t}\}_{t=1}^{2000}$ of the model in Equation \eqref{eq:scaleStudUniMeas}, \eqref{eq:scaleStudUniTrans}, filtered estimates computed using Equation \eqref{eq:pred}, and filtered estimates computed using the QMLE approach of \cite{RUIZ1994289} based on standard Kalman filtering. Right:  predictive state variance $\{p_t\}_{t=1}^{2000}$ computed using Equation \eqref{eq:pred}, and predictive state variance computed using the standard Kalman filter.   }
\label{fig:exampleFilt}
\end{figure}

The robustness effect provided by the score and Hessian functions is also described in Figure \eqref{fig:exampleFilt}, which shows the results of one simulation of the model in Equation \eqref{eq:scaleStudUniMeas}, \eqref{eq:scaleStudUniTrans} with $\nu=3$. We report on the left the simulated scale trajectory $\{e^{\alpha_t}\}_{t=1}^{2000}$, together with the predictive estimates computed using Equation \eqref{eq:pred}, and those computed using the QMLE approach of \cite{RUIZ1994289}. The latter approach applies the standard Kalman filter to the linearized measurement equation $\log y_t^2=\alpha_t+ \log(\epsilon_t^2)$ assuming a normal distribution for $\log(\epsilon_t^2)$. On the right, we report the conditional variance $p_t=\mathbb{V}[\alpha_t|Y_{t-1}]$ computed using Equation \eqref{eq:pred} and the standard Kalman filter.\ As can be seen from the left figure, the filtered estimates of the Kalman filter are significantly affected by non-normality because they overreact to the extremely large observations generated by the Student-$t$ innovations in Equation \eqref{eq:scaleStudUniMeas}. The right figure shows that this lack of robustness leads to a spuriously large state variance. In fact, the Kalman filter provides a variance $p_t$ significantly larger than the state variance computed using the robust recursions in Equation \eqref{eq:pred}. Note also that, contrary to the Kalman filter variance, the variance computed using our approach varies stochastically over time. This property breaks the steady-state and leads to non-trivial dynamics for $p_t$ assigning different weights to different observations. The absence of a steady-state is commonly observed in nonlinear and non-Gaussian models. 

\subsection{Comparison with related filtering methods}

The winsorizing effect on the mean and variance estimates is common to other filtering methods based on the score and the Hessian matrix, for example the methods of \cite{Masreliez} and \cite{West81}.\
Although such methods employ the score and Hessian functions to update the first two conditional moments, they lead to different recursive estimates.\ In the method of \cite{Masreliez}, the score and the Hessian are not available in closed form, and must be approximated using numerical convolutions.\ \cite{West81} simplifies the \cite{Masreliez} filter using a heuristic approach. Both methods are derived assuming a location model as a data generating process and a Gaussian prior density for the latent states. We instead consider the general state-space model in Equation \eqref{eq:ssm_obs_gen}, \eqref{eq:ssm_trans_gen} encompassing not only location, but also several other types of state-space models.\ Furthermore, as discussed in Section \eqref{sub:approxResults}, our assumptions on the tail behavior of the prior density are significantly milder compared to normality. Relaxing the normality assumption is important because the prior density is clearly non-normal in nonlinear and non-Gaussian state-space models.\ \cite{SchickMitter94} avoid this assumption by representing the prior density as a member of the $\epsilon$-contaminated normal neighborhood.\ However, this comes at the cost of significantly higher computational complexity. Their method involves a bank of filtering and smoothing recursions running in parallel having a more complex expression for the score function depending on the particular choice of the contaminating distribution.\ Moreover, as underlined by the authors, choosing the contaminating distribution is a problem whose solution remains incomplete. 

Another filtering methodology based on the score function is the approximation via mode estimation described by \cite{DurbinKoopman}; see also \cite{DurbinKoopman2000} and \cite{ShephardPitt}.\ This method is an optimization algorithm approximating the mode of the filtering density. As shown in Section \eqref{sub:approxResults}, we instead approximate the mean and variance of the filtering density.\ Moreover, the approximation via mode estimation embeds the classical Kalman filter recursions (running in time) into Newton-Raphson recursions (running in the state space). This leads to a recursive structure different from the one presented above, which only involves filtering recursions running in time. 

The presence of the score of the observation density as a driving force for the filter makes our methodology comparable to the class of score-driven time-series models of \cite{GAS1} and \cite{Harvey_2013}; see also \cite{OhPatton}, \cite{LucasEtAl2019}, \cite{harvey2023score} for some recent applications
of this class of models.\ Score-driven models are purely predictive time-series models where the time-varying parameters are completely revealed by past information.\ The filtered estimates provided by score-driven models carry no filtering uncertainty because conditioning on past information is sufficient to remove all of the uncertainty.\ In our framework, the latent states are not pre-determined given past observations, and therefore their conditional distribution is non-degenerate.\ As a consequence, contemporaneous and subsequent information can be used in order to reduce filtering uncertainty and improve the estimation of the latent states.\ This can be done through the updating and smoothing steps, which we derive together with the classical predictive step. Our framework also introduces parallel recursions for the conditional variance characterizing the dispersion of filtering uncertainty. The methodology can thus be regarded as an extension of the classical score-driven framework allowing for non-degenerate filtering distributions and the computation of update and smoothed estimates.

\section{Derivation of the recursions}
\label{sec:statProp}


\subsection{The perturbation approach}
\label{sub:intuitive}

We first provide the intuition behind the method used to derive the approximate filtering recursions described in Section \eqref{sec:filtSmooth}. We restrict our attention to the following univariate specification of the state-space model in Equation \eqref{eq:ssm_obs_gen}, \eqref{eq:ssm_trans_gen}:\medskip 
\begin{align}
 {y}_t|{\alpha}_t & \sim p({y}_t|{\alpha}_t;\theta)\label{eq:ssm_obs}\\
 {\alpha}_{t+1} &= {c} + \phi{\alpha}_t + {\eta}_t\label{eq:ssm_trans}
 \end{align}
where $|\phi|<1$ and $\eta_t$ is a sequence of i.i.d.\ innovations with zero mean and variance $q$. 
The extension of our results to a multivariate setting can be undertaken using the same approach at the expense of more complex notations.  

Let us denote by $p(\alpha_t|Y_{t-1})$ (resp. $p(\alpha_t|Y_{t})$) the conditional density of $\alpha_t$ given observations up to time $t-1$ (resp. $t$).\ We refer to $p(\alpha_t|Y_{t-1})$ as the \textit{prior} distribution of $\alpha_t$.\  For ease of notation, we write $p(y_t|\alpha_t)$ in place of $p(y_t|\alpha_t;\theta)$.
We are interested in finding an approximation to the following integral:\medskip
\begin{equation}
\mathcal{G}_t^{(\ell)}(y_1,\dots,y_{t})=N_t\int_{-\infty}^{+\infty}\alpha_t^{\ell} p(y_t|\alpha_t)p(\alpha_t|Y_{t-1})d\alpha_t
\label{eq:intuitive}\medskip
\end{equation} 
where $\ell\in\mathbb{N}_0$ and $N_t$ is a normalizing coefficient.\ For example, if $\ell=0$ and $N_t=1$ for any $t$, the function $\mathcal{G}_t^{(\ell)}(y_1,\dots,y_{t})$ corresponds to the conditional likelihood $p(y_t|Y_{t-1})$; if instead $\ell>0$ and $N_t=p(y_t|Y_{t-1})^{-1}$, $\mathcal{G}_t^{(\ell)}(y_1,\dots,y_{t})$ reduces to the $\ell$-th moment of $p(\alpha_t|Y_t)$. 

For nonlinear and non-Gaussian models, $\mathcal{G}_t^{(\ell)}(y_1,\dots,y_{t})$ cannot be computed in closed form.\ Some approximation methods simplify the integral by assuming that $p(\alpha_t|Y_{t-1})$ is Gaussian (\citealt{Masreliez}, \cite{Martin79}, \citealt{West81}); others use Gaussian sum approximations, such as \cite{SORENSON1971465}, or Gram–Charlier and Edgeworth expansions, such as \cite{sorenson1988}.

The method we adopt here builds upon the observation that, if $p(y_t|\alpha_t)$ has a polynomial representation in terms of powers of $\alpha_t$, the integral of $\alpha_t^{\ell}p(y_t|\alpha_t)$ with respect to the measure $p(\alpha_t|Y_{t-1})d\alpha_t$ reduces to a sum of past moments of the prior density. For example, let us set $\ell=1$ and assume $p(y_t|\alpha_t)=\omega_0 + \omega_1\alpha_t$, where the coefficients $\omega_0$, $\omega_1$ could depend on $y_t$. Neglecting the normalizing coefficient, the integral in Equation \eqref{eq:intuitive} becomes: \bigskip
\begin{align}
\int_{-\infty}^{+\infty}\alpha_t^{\ell} p(y_t|\alpha_t)p(\alpha_t|Y_{t-1})d\alpha_t=\int_{-\infty}^{+\infty} (\omega_0\alpha_t + \omega_1\alpha_t^2)p(\alpha_t|Y_{t-1})d\alpha_t=\omega_0 a_t + \omega_1(p_t+a_t^2) \medskip
\label{eq:exampleInt}
\end{align}
where $a_t$ and $p_t$ denote the mean and variance of $p(\alpha_t|Y_{t-1})$. Given the Markovian structure of the transition equation, $a_t$, $p_t$ can be written in terms of $a_{t-1|t-1}$, $p_{t-1|t-1}$, which denote the mean and variance of $p(\alpha_{t-1}|Y_{t-1})$.\ The latter two quantities can in turn be computed by evaluating $\mathcal{G}_{t-1}^{(\ell)}(y_1,\dots,y_{t-1})$ as in Equation \eqref{eq:exampleInt}.\ This leads to a recursive algorithm expressing the conditional moments of the latent states as a function of past conditional moments. 

Since $p(y_t|\alpha_t)$ is not generally a polynomial in $\alpha_t$, we expand the observation density in a power series in order to obtain a polynomial representation that can be integrated with respect to $p(\alpha_t|Y_{t-1})d\alpha_t$.\ That is, we write the integral in Equation \eqref{eq:intuitive} as follows:\bigskip
\begin{equation}
\mathcal{G}_t^{(\ell)}(y_1,\dots,y_{t})=N_t\int_{-\infty}^{+\infty}\alpha_t^{\ell} [p^{[k]}(y_t|\alpha_t)+ g(\alpha_t)]p(\alpha_t|Y_{t-1})d\alpha_t
\label{eq:intuitivePower}\medskip
\end{equation} 
where $p^{[k]}(y_t|\alpha_t)$ is the $k$-th order power series expansion of $p(y_t|\alpha_t)$ and $g(\alpha_t)$ is the remainder of the expansion. The integral of $\alpha_t^{\ell} p^{[k]}(y_t|\alpha_t)$ with respect to $p(\alpha_t|Y_{t-1})d\alpha_t$, which can be computed in closed form, coincides with the true $\ell$-th moment of $p(\alpha_t|Y_t)$ up to an error depending on the remainder $g(\alpha_t)$.\ We show below that this approximate moment can be computed recursively under a set of tail conditions on the prior density $p(\alpha_t|Y_{t-1})$ guaranteeing the convergence of the integral in Equation \eqref{eq:intuitivePower}.\ In particular, such conditions ensure the finiteness of the moments of $p(\alpha_t|Y_{t-1})$ up to the order $\ell+k$, so that the approximate $\ell$-th moment  $\int_{-\infty}^{+\infty}\alpha_t^{\ell} p^{[k]}(y_t|\alpha_t)p(\alpha_t|Y_{t-1})d\alpha_t$ is finite, and also the finiteness of the reminder term $\int_{-\infty}^{+\infty}\alpha_t^{\ell}g(\alpha_t)p(\alpha_t|Y_{t-1})d\alpha_t$.\ Since such tail conditions are significantly weaker compared to the normality assumption of \cite{Masreliez}, \cite{Martin79} and \cite{West81}, our results show that the use of the score and Hessian functions is robust for a wide class of state-space models and in realistic scenarios where the prior density deviates from normality. 


Since the expansion of $p(y_t|\alpha_t)$ is around the predictive filter $a_t$, the approximation error is small when there is a large probability that $\alpha_t$ is close to $a_t$, i.e., when $p(\alpha_t|Y_{t-1})$ tends to a degenerate density. We thus obtain the approximate general solution to the filtering problem in a neighborhood of the particular solution corresponding to a degenerate prior density. For example, if the variance $q$ of the state innovations is very small, $\alpha_t$ does not deviate significantly from its predictive estimate, and thus we expect the approximate filter performing closely to the true optimal filter. On the contrary, when $q$ is larger, we expect a deterioration of the approximate filter with respect to the true optimal filter. We illustrate this phenomenon in the Monte Carlo analysis of Section \eqref{sec:MonteCarlo}. 


\subsection{Approximation results}
\label{sub:approxResults}

In this section, we derive a number of approximation results expressing recursively the conditional moments of the latent states. These results are not immediately applicable to obtain feasible filtering and smoothing recursions because they approximate the true optimal estimates at time $t$ in terms of the true optimal estimates at time $t-1$, which are unknown. 
We discuss at the end of the section how a set of feasible recursions can be recovered from such results. 
In the following, $\beta_t^{(j)}$, $j>1$, denote the $j$-th absolute central moments of $p(\alpha_t|Y_{t-1})$, whereas $p_t^{(j)}$, $j>2$, denote the $j$-th central moments of $p(\alpha_t|Y_{t-1})$.  We first study the case $\ell=0$ and $N_t=1$ in Equation \eqref{eq:exampleInt}, which provides an approximation to the conditional likelihood $p(y_t|Y_{t-1})$.

\begin{assump}
The observation density $p(y_t|\alpha_t)$ is a bounded function of $\alpha_t$, twice differentiable in $\mathbb{R}$ with continuous second derivative. 
\label{ass:1}
\end{assump}

\begin{assump}
For any $t\in\mathbb{N}$, there exist $\bar{\delta}>0$ and $0<\beta<1$ such that $p(\alpha_t|Y_{t-1})\le \frac{1}{|\alpha_t-a_t|^{3+\beta}}$ for all $\alpha_t$ such that $|\alpha_t-a_t|\ge \bar{\delta}$.
\label{ass:2} 
\end{assump}

\noindent Assumption \eqref{ass:1} implicitly requires that $p(y_t|\alpha_t)$ is defined for any $\alpha_t\in\mathbb{R}$. This assumption is not restrictive because, if $\alpha_t\in \mathcal{C}\subset\mathbb{R}$ is bounded, it is always possible to find a differentiable link function $f:\mathcal{C}\to\mathbb{R}$ transforming $\alpha_t$ into a new state variable defined in $\mathbb{R}$. The condition in Assumption \eqref{ass:2} allows the prior density being non-normal and also asymmetric, but restricts its tails to decay faster than the third power of $\alpha_t-a_t$. It also implies the existence and finiteness of the variance of $p(\alpha_t|Y_{t-1})$.  As shown in Appendix \eqref{app:prop:pyt}, the following result holds:

\begin{prop}
 Let Assumptions \eqref{ass:1},\eqref{ass:2} hold. Then, for any $\gamma>0$, we can find $\tilde{\delta}\ge\bar{\delta}$ such that:
 \begin{equation}
  p(y_t|Y_{t-1}) = p(y_t|\alpha_t=a_t) + \xi_t\medskip
  \label{eq:prop:pyt}
 \end{equation}
with $\displaystyle |\xi_t| \le \gamma+\frac{1}{2}\sup_{|\alpha_t-a_t|\le \tilde{\delta}}\left |\frac{\partial^2 p(y_t|\alpha_t)}{\partial\alpha_t^2}\right |p_t$.
\label{prop:pyt}\medskip
\end{prop}

\noindent This result characterizes the error arising when approximating the conditional likelihood $p(y_t|Y_{t-1})$ with the observation density $p(y_t|\alpha_t)$ evaluated for $\alpha=a_t$. 
The error includes a component, $\gamma$, that can be made arbitrarily small, and an additional term whose magnitude depends on $p_t$, the variance of $p(\alpha_t|Y_{t-1})$.\ Since $p_t$ is finite, the error is bounded and tends to be small when $p(\alpha_t|Y_{t-1})$ is close to a degenerate density.\ 


We now set $\ell=1$ and $N_t=p(y_t|Y_{t-1})^{-1}$ in Equation \eqref{eq:intuitivePower} in order to approximate the conditional mean of $p(\alpha_t|Y_{t})$. Compared to the case $\ell=0$, the degree of the polynomial appearing inside the integral in Equation \eqref{eq:intuitivePower} increases by one. To guarantee the boundedness of the approximation error, a more restrictive condition on the tails of $p(\alpha_t|Y_{t-1})$ is needed. 

\begin{assump}
For any $t\in\mathbb{N}$, there exist $\bar{\delta}>0$ and $0<\beta<1$ such that $p(\alpha_t|Y_{t-1})\le \frac{1}{|\alpha_t-a_t|^{4+\beta}}$ for all $\alpha_t$ such that $|\alpha_t-a_t|\ge \bar{\delta}$.
\label{ass:3} 
\end{assump}
\noindent This assumption implies the existence and finiteness of the moments of $p(\alpha_t|Y_{t-1})$ up to the third order. Note also that Assumption \eqref{ass:3} implies Assumption \eqref{ass:2}, so that, if Assumptions \eqref{ass:1} and \eqref{ass:3} are satisfied, Proposition \eqref{prop:pyt} holds. 
In Appendix \eqref{app:prop:att}, we prove the following theorem:
\begin{theorem}
Let Assumptions \eqref{ass:1},\eqref{ass:3} hold. Then, for any $\gamma>0$, we can find $\tilde{\delta}\ge \bar{\delta}$ such that:
 \begin{equation}
  a_{t|t} = a_t + p_t\nabla_t(a_t) + \chi_t + O(\xi_t) \medskip
  \label{eq:prop:att1}
 \end{equation}
where $\displaystyle   |\chi_t|\le \gamma +\frac{1}{2p(y_t| \alpha_t)|_{a_t}}\sup_{|\alpha_t-a_t|\le \tilde{\delta}}\left |\frac{\partial^2 p(y_t|\alpha_t)}{\partial\alpha_t^2}\right |\left(\beta_t^{(3)}+|a_t|p_t\right)$ and $O(\xi_t)$ denotes terms of order $\xi_t$. 
\label{prop:att}
\end{theorem}
\noindent The approximate update filter $a_t + p_t\nabla_t(a_t)$ differs from the true optimal update filter $a_{t|t}$ by a bounded error term $\chi_t$ and other terms of order $\xi_t$; see Proposition \eqref{prop:pyt}. In particular, the term $\chi_t$, which arises upon integrating the reminder in the power series expansion of $p(y_t|\alpha_t)$, is small when the prior density tends to a degenerate density.\ Note that its magnitude is related not only to the variance, but also to the third absolute central moment of $p(\alpha_t|Y_{t-1})$. 
Combining the result in Theorem \eqref{prop:att} with Equation \eqref{eq:ssm_trans}, it is simple to obtain a similar approximation result for the predictive filter $a_{t+1}$. Indeed, under the same assumptions of Theorem \eqref{prop:att}, we have:
\begin{equation}
  a_{t+1} = c + \phi a_t + \phi p_t\nabla_t(a_t)  + \phi\chi_t + O(\xi_t) 
  \label{eq:prop:at1}
 \end{equation}
 which relates the approximate predictive filter $c + \phi a_t + \phi p_t\nabla_t(a_t)$ to the true predictive filter $a_{t+1}$. 

Both Proposition \eqref{prop:pyt} and Theorem \eqref{prop:att} are recovered by considering a first-order expansion of $p(y_t|\alpha_t)$, i.e., by setting $k=1$ in Equation \eqref{eq:intuitivePower}. In order to obtain an approximation result for the two conditional variances $p_{t|t}$ and $p_{t+1}$, we perform a second-order expansion, i.e., we set $\ell=2$ and $k=2$ in Equation \eqref{eq:intuitivePower}. This requires a higher degree of smoothness for $p(y_t|\alpha_t)$ and a stricter condition for the tail behavior of the prior density.


\begin{assump}
The observation density $p(y_t|\alpha_t)$ is a bounded function of $\alpha_t$, three times differentiable in $\mathbb{R}$ with continuous third derivative. 
\label{ass:4}
\end{assump}

\begin{assump}
For any $t\in\mathbb{N}$, there exist $\bar{\delta}>0$ and $0<\beta<1$ such that $p(\alpha_t|Y_{t-1})\le \frac{1}{|\alpha_t-a_t|^{6+\beta}}$ for all $\alpha_t$ such that $|\alpha_t-a_t|\ge \bar{\delta}$.
\label{ass:5} 
\end{assump}
\noindent Assumption \eqref{ass:5} implies now the existence and finiteness of the moments of $p(\alpha_t|Y_{t-1})$ up to the fifth order.
The following result is proved in Appendix \eqref{app:prop:ptt}:
\begin{theorem}
Let Assumptions \eqref{ass:4},\eqref{ass:5} hold. Then, for any $\gamma>0$, we can find $\tilde{\delta}\ge \bar{\delta}$ such that:
 \begin{equation}
  p_{t|t} =  p_t - p_t^2\nabla_t^2(a_t)+ \ell_t + \zeta_t + O(\xi_t) + O(\chi_t) \medskip
  \label{eq:prop:ptt1}
 \end{equation}
where $\ell_t= \nabla_t(a_t) p_t^{(3)} + \frac{1}{2p(y_t|\alpha_t)|_{a_t}}\frac{\partial^2 p(y_t|\alpha_t)}{\partial\alpha_t^2}\bigg |_{a_t}p_t^{(4)}$,  $|\zeta_t|\le \gamma + \frac{1}{6p(y_t|\alpha_t)|_{a_t}}\sup_{|\alpha_t-a_t|\le \tilde{\delta}} \left| \frac{\partial^3 p(y_t|\alpha_t)}{\partial\alpha_t^3}\right|\beta_t^{(5)}$ and $O(\xi_t)$, $O(\chi_t)$ denote terms of order $\xi_t$, $\chi_t$, respectively.
\label{prop:ptt}
\end{theorem}
\noindent 
This theorem shows that  $p_t-p_t^2 \nabla_t^2(a_t)$ is an approximation to the conditional variance $p_{t|t}$. The terms in $\zeta_t$, $\xi_t$, $\chi_t$ are bounded errors resulting from the expansion and subsequent integration of the reminder in Equation \eqref{eq:intuitivePower}. The term $\ell_t$ depends on the third and fourth conditional moments of $p(\alpha_t|Y_{t-1})$. 
It is interesting to note that, when $p(\alpha_t|Y_{t-1})$ is normal or fat-tailed, a more accurate approximation to $p_{t|t}$ is given by $p_t+p_t^2 h_t(a_t)$, where $h_t(a_t)=\frac{\partial^2\log p(y_t|\alpha_t)}{\partial \alpha_t^2}\bigg |_{a_t}$ denotes the second derivative of $\log p(y_t|\alpha_t)$ evaluated for $\alpha_t=a_t$. Indeed, the following corollary of Theorem \eqref{prop:ptt} holds: 
\begin{corol}
Let Assumptions \eqref{ass:4},\eqref{ass:5} hold. Then, for any $\gamma>0$, we can find $\tilde{\delta}\ge \bar{\delta}$ such that:
 \begin{equation}
  p_{t|t} =  p_t + p_t^2h_t(a_t)+ \ell_t' + \zeta_t + O(\xi_t) + O(\chi_t) \medskip
  \label{eq:corol:ptt1}
 \end{equation}
where $\ell_t'= \nabla_t(a_t) p_t^{(3)} + \frac{1}{2p(y_t|\alpha_t)|_{a_t}}\frac{\partial^2 p(y_t|\alpha_t)}{\partial\alpha_t^2}\bigg |_{a_t}\left(p_t^{(4)} - 2p_t^2\right)$,  $|\zeta_t|\le \gamma + \frac{1}{6p(y_t|\alpha_t)|_{a_t}}\sup_{|\alpha_t-a_t|<\tilde{\delta}} \left| \frac{\partial^3 p(y_t|\alpha_t)}{\partial\alpha_t^3}\right|\beta_t^{(5)}$ and $O(\xi_t)$, $O(\chi_t)$ denote terms of order $\xi_t$, $\chi_t$, respectively.
\label{corol:ptt}
\end{corol}
\noindent This result is an immediate consequence of the following relation involving the second derivative of $\log p(y_t|\alpha_t)$ and the square of the score:
\begin{align*}
&\quad \frac{1}{p(y_t|\alpha_t)|_{a_t}}\frac{\partial^2 p(y_t|\alpha_t)}{\partial\alpha_t^2}\bigg |_{a_t} - \nabla_t^2(a_t)\\
& = \frac{1}{p(y_t|\alpha_t)|_{a_t}}\frac{\partial^2 p(y_t|\alpha_t)}{\partial\alpha_t^2}\bigg |_{a_t} - \left(\frac{1}{p(y_t|\alpha_t)|_{a_t}}\frac{\partial p(y_t|\alpha_t)}{\partial\alpha_t}\bigg |_{a_t}\right)^2\\
& = \frac{\partial^2\log p(y_t|\alpha_t)}{\partial \alpha_t^2}\bigg |_{a_t}
\end{align*}
Since $0\le p_t^{(4)}-2p_t^2\le p_t^{(4)}$ for normal and fat-tailed distributions\footnote{With the term fat-tailed distribution, we refer here to a probability distribution having excess kurtosis larger than zero.}, we have $|\ell_t'|\le |\ell_t|$, meaning that the approximation error in Corollary \eqref{corol:ptt} is smaller than the one in Theorem \eqref{prop:ptt}. Observe also that the result in Theorem \eqref{prop:ptt} can be combined with Equation \eqref{eq:ssm_trans} to obtain an approximation for $p_{t+1}$. Indeed, under the same assumptions of Theorem \eqref{prop:ptt}, we have:
 \begin{equation}
  p_{t+1} =  \phi^2(p_t - p_t^2\nabla_t^2(a_t))+ q + \phi^2\ell_t + \phi^2\zeta_t + O(\xi_t) + O(\chi_t) 
  \label{eq:prop:pt1}
 \end{equation}
which characterizes the error obtained by approximating $p_{t+1}$ with $\phi^2(p_t - p_t^2\nabla_t^2(a_t))+ q$. A similar approximation result in terms of $h_t(a_t)$ can also be derived.

\begin{remark}
When the model is linear and Gaussian, i.e., $p(y_t|\alpha_t)=N(\alpha_t,r)$, $r\in\mathbb{R}^+$, we obtain that the score $\nabla_t(a_t)=\frac{y_t-a_t}{r}$ is linear and the Hessian $h_t(a_t)=-\frac{1}{r}$ is independent from the sample data, as in the Kalman filter. However, the approximation error in Theorems \eqref{prop:att}, \eqref{prop:ptt} and in Corollary \eqref{corol:ptt} is nonzero. In fact, the filter we obtain is not exactly equal to the Kalman filter under the true model parameters, but coincides with it when the parameters are estimated using the approximate log-likelihood $\sum_{t=1}^n\log p(y_t|\alpha_t=a_t)=\sum_{t=1}^n -\frac{1}{2}\left[\log(2\pi r)+\frac{(y_t-a_t)^2}{r}\right]$. This happens because the center around which we perturb the filtering density does not correspond to a Gaussian prior density, but to a degenerate prior density.\ Consequently, the approximation error is zero only when $p(\alpha_t|Y_{t-1})$ is a degenerate density.
\label{remark:KF}
\end{remark}

The previous results show that the second conditional moments depend on the past third and fourth conditional moments. More generally, if we approximate the conditional moments up to the $\ell$-th order, the $\ell$-th moment will depend on the past $(\ell+1)$-th moment if a first-order expansion is considered (as in Theorem \ref{prop:att}), and on the past $(\ell+1)$ and $(\ell+2)$-th moments if a second-order expansion is performed (as in Theorem \ref{prop:ptt}). Similarly, the magnitude of the error arising by integrating the reminder will depend on the $(\ell+2)$-th absolute central moment in the case of a first-order expansion, and on the $(\ell+3)$-th absolute central moment in the case of a second-order expansion. Clearly, approximating such higher order moments requires stricter tail conditions for the prior density.

We finally discuss how similar approximation results can be obtained for the smoothed estimates. The main difference with respect to the filtered estimates is that analogous tail conditions for the transition density $p(\alpha_t|\alpha_{t-1})$ are needed. This can be seen from the following result proved in Appendix \eqref{app:prop:smoothG}: 
\begin{prop}
The moments of the smoothing density $p(\alpha_t|Y_n)$, $t\le n$, can be written as follows:
\begin{equation}
\mathcal{S}_t^{(\ell)}(y_1,\dots,y_n)=N_t'\int_{-\infty}^{+\infty}\alpha_t^{\ell}K_t(\alpha_t) p(y_{t}|\alpha_{t})p(\alpha_t|Y_{t-1})d\alpha_t \medskip
\label{eq:smoothS}
\end{equation}
where $\ell\in\mathbb{N}$, $N_t'=p(y_t,\dots,y_n|Y_{t-1})^{-1}$ and $K_t(\alpha_t)$ is defined recursively as follows:
\begin{equation*}
 K_i(\alpha_i)=\int_{-\infty}^{+\infty} K_{i+1}(\alpha_{i+1})p(y_{i+1}|\alpha_{i+1})p(\alpha_{i+1}|\alpha_i)d\alpha_{i+1}
\end{equation*} 
for $i=t,t+1,\dots,n-1$, and $K_{n}=1$.
\label{prop:smooth}\medskip
\end{prop}
\noindent The integral in Equation \eqref{eq:smoothS} defining the smoothed estimates has a structure similar to the one in Equation \eqref{eq:intuitive}, but includes the recursive kernel $K_t(\alpha_t)$. The role of this kernel is to account for the effect of the observations from time $t+1$ up to time $n$ in the estimation of $\alpha_t$. Thus, to compute the integral, it is first necessary to approximate $K_i(\alpha_i)$, for $i=n-1,\dots,t$. This can be done as above, i.e., expanding $p(y_{i+1}|\alpha_{i+1})$ around $a_{i+1}$ and integrating with respect to $p(\alpha_{i+1}|\alpha_i)d\alpha_{i+1}$. To ensure the finiteness of such integrals, it is necessary to impose a tail condition on $p(\alpha_{i+1}|\alpha_i)$ guaranteeing the existence of its moments up to a certain order and the boundedness of the approximation error. This procedure gives rise to a backward recursion approximating the conditional moments of $\alpha_t$. For example, for $\ell =1$, integrating with respect to $p(\alpha_{i+1}|\alpha_i)d\alpha_{i+1}$, for $i=n-1,\dots,t$, and finally with respect to $p(\alpha_t|Y_{t-1})d\alpha_t$, leads to a weighted sum of scores from $t+1$ to $n$ with weights decaying as $\phi^j$, $j=1,\dots, n-t$, similar to the scaling implied by the recursions in Equation \eqref{eq:smooth}.

The results recovered above approximate the (true) conditional moments in terms of past (true) conditional moments, which are unknown. This is common to other approximate filtering techniques based on local or global expansions, for example the Extended Kalman filter or the methods of \cite{Masreliez} and \cite{West81}.\ Moreover, each conditional moment depends on past higher order conditional moments. For example, $a_{t+1}$ depends on $p_t$, which in turn depends on $p_{t-1}^{(3)}$ and $p_{t-1}^{(4)}$. Therefore, in order to use these results in practice, it is necessary to replace the past true conditional moments with the past approximate conditional moments and retain the moments up to a certain order in the expansion. We keep the terms up to the second moments in our approximation results. This leads to a set of feasible forward-backward recursions identical to those in Section \eqref{sec:filtSmooth} which can be run starting from two initial values ${a}_{1}$, ${p}_{1}$ for the predictive mean and variance. We set such initial values equal to the unconditional mean and variance of $\alpha_t$, as in the standard Kalman filter. In principle, the third and fourth moments could be kept and approximated using the same method. This would provide a more complete characterization of the filtering distribution, but would also lead to more complex recursions and stricter assumptions on the prior density.  

The presence of the Hessian in place of the squared score in the recursions of Section \eqref{sec:filtSmooth} is motivated by the result in Corollary \eqref{corol:ptt}. In fact, the Hessian reduces the approximation error and provides better estimates of the conditional variance.\  The conditional variance estimates obtained with our recursions are not positive-definite with probability one.\ Also this aspect is common to other filtering methods based on local or global expansions, such as \cite{West81} or the Gram-Charlier and Edgeworth expansions in \cite{sorenson1988}.\ A non positive-definite variance estimate is more likely to occur in Gaussian scale models where the Hessian becomes large and negative when $|y_t|$ is large; see right panel of Figure \eqref{fig:hessianScore}.\ When this happens, one possible method to obtain positive-definite variance estimates is to truncate the Hessian for very large negative values, as in \cite{West81}, or to set the covariance matrix $\mb{P}_t$ equal to zero or to a diagonal matrix with very small diagonal entries.\ We use the latter method in our simulation and empirical analysis below.\ It is worth pointing out that, in our Monte Carlo experiments, we obtain only a few cases (lower than 3\%) in which the covariance is not positive definite, all pertaining the Gaussian scale model. 

\subsection{Parameter and filtering uncertainty}
\label{sub:filtIUnc}


Given the conditional mean and variance of the latent states, it is possible to compute approximate confidence bands around the filtered and smoothed estimates.\ For example, an approximation to the quantiles of $p(\boldsymbol{\alpha}_t|\mathbf{Y}_{t-1})$, $p(\boldsymbol{\alpha}_t|\mb{Y}_{t})$, $p(\boldsymbol{\alpha}_t|\mb{Y}_{n})$ can be computed using the Chebyshev's inequality or assuming a Gaussian distribution with mean and variance given by the conditional estimates.\ More accurate results taking into account the non-normality of $p(\boldsymbol{\alpha}_t|\mathbf{Y}_{t-1})$ and $p(\boldsymbol{\alpha}_t|\mb{Y}_{t})$ can be computed using a Student-$t$ distribution whose number of degrees of freedom is selected by cross-validation.

These confidence bands reflect filtering uncertainty, but neglect the uncertainty arising from parameters estimation.\ Parameter uncertainty refers to the fact that the static parameters of the state-space model are unknown and must be estimated from the data. 
In order to account for the additional uncertainty surrounding the maximum likelihood estimates, we follow the same approach employed in linear and Gaussian models. In particular, we adopt the Bayesian perspective that the static parameters vector $\bs{\theta}$ is a random variable with a certain prior distribution $p(\bs{\theta})$.  Let $\mb{a}_t^{\bs{\hat{\theta}}}$ denote the predictive filter computed using the estimate $\bs{\hat{\theta}}$. The total conditional variance of the latent states can be written as the sum of two terms (see \citealt{HAMILTON1986387}):
\begin{equation}
\begin{split}
 \mathbb{E}(\bs{\alpha}_t-\mb{a}_t^{\bs{\hat{\theta}}})(\bs{\alpha}_t-\mb{a}_t^{\bs{\hat{\theta}}})'|\mb{Y}_{t-1}] &=\\
 \mathbb{E}_{{\bs{\theta}}}[(\bs{\alpha}_t-\mb{a}_t^{\bs{{\theta}}})(\bs{\alpha}_t- \mb{a}_t^{\bs{{\theta}}})'|\mb{Y}_{t-1}] &+ \mathbb{E}_{{\bs{\theta}}}[(\mb{a}_t^{\bs{\theta}}-\mb{a}_t^{\bs{\hat{\theta}}})(\mb{a}_t^{\bs{\theta}}-\mb{a}_t^{\bs{\hat{\theta}}})']=\\
 \mathbb{E}_{\bs{\theta}}[\mb{P}_t^{\bs{\theta}}] &+ \mathbb{E}_{\bs{\theta}}[(\mb{a}_t^{\bs{\theta}}-\mb{a}_t^{\bs{\hat{\theta}}})(\mb{a}_t^{\bs{\theta}}-\mb{a}_t^{\bs{\hat{\theta}}})']
 \end{split}
 \label{eq:filtParUnc}
\end{equation}
where $\mathbb{E}_{\bs{\theta}}[\cdot]$ denotes the expectation with respect to the prior density $p(\bs{\theta})$. The first term is related to filtering uncertainty because it represents the average conditional variance of the latent states. The second term is instead related to parameter uncertainty because it represents the variation of $\mb{a}_t^{\bs{\theta}}$ imputable to the randomness of $\bs{\theta}$. Both terms can be evaluated by Monte Carlo simulations, sampling from the prior density $p(\bs{{\theta}})$. In practical applications, $p(\bs{\theta})$ is set equal to the asymptotic distribution of the maximum likelihood estimate $\bs{\hat{\theta}}$, which is assumed to be normal with variance given by the inverse Fisher information. Since the variance decomposition in Equation \eqref{eq:filtParUnc} also holds when conditioning with respect to $\mb{Y}_t$ and $\mb{Y}_n$, the same methodology can be adopted to compute confidence bands around the update and smoothed estimates. 


\section{Monte Carlo analysis}
\label{sec:MonteCarlo}

In this section we examine by Monte Carlo simulations the performance of the approximate filtering and smoothing methodology presented in Section \eqref{sec:filtSmooth}.\ The goal is to assess the mean square loss incurred by the local approximation through a comparison with exact simulation-based methods. As a data generating process, we consider four nonlinear and/or non-Gaussian state-space models having the same form of Equation \eqref{eq:ssm_obs_gen}, \eqref{eq:ssm_trans_gen}.  Their measurement equations are given below:
\begin{align*}
\text{Location (Student-\textit{t}):}\quad \quad \quad y_t &= \alpha_t + \epsilon_t,\quad \epsilon_t\sim t_{\nu}(0,e^{\lambda})\\
\text{Scale (Gaussian):}\quad \quad\quad y_t &= e^{\frac{\alpha_t}{2}}\epsilon_t,\quad \epsilon_t \sim \text{N}(0,1)\\
\text{Scale (Student-\textit{t}):}\quad\quad \quad y_t &= e^{\frac{\alpha_t}{2}}\epsilon_t,\quad \epsilon_t \sim t_{\nu}(0,1)\\
\text{Count data (Poisson):}\quad \quad\quad y_t &\sim \text{Poiss}(\alpha_t)
\end{align*}
where the state variable $\alpha_t$ evolves as follows:
\begin{equation}
\alpha_{t+1} = {c} + \phi{\alpha}_t + {\eta}_t,\quad \eta_t\sim \text{N}(0,q)
\label{eq:ssm_trans_gauss}
\end{equation}


\begin{table}[htb]
\centering
\begin{adjustbox}{max width=1.1\textwidth,center}
\begin{tabular}{c|c|c|c}
  \hline
 Model   & $p(y_t|\alpha_t=a_t)$ & $\nabla_t(a_t)$ & $h(a_t)$\\
  \hline 
   Location    & $\frac{\Gamma(\frac{\nu+1}{2})}{\Gamma(\frac{\nu}{2})\sqrt{\pi(\nu-2)e^{\lambda}}}\left[1+\frac{(y_t-a_t)^2}{(\nu-2) e^{\lambda}}\right]^{-\frac{\nu+1}{2}}$&  $(\nu+1)\frac{y_t-a_t}{(\nu-2)e^{\lambda}+(y_t-a_t)^2}$ & $(\nu+1)\frac{(y_t-a_t)^2-(\nu-2)e^{\lambda}}{[(\nu-2)e^{\lambda}+(y_t-a_t)^2]^2}$\\
   Scale & $\frac{1}{\sqrt{2\pi e^{a_t}}}\exp(-\frac{y_t^2}{2e^{a_t}})$ & $\frac{1}{2}\left[\frac{y_t^2}{e^{a_t}}-1 \right]$ & $-\frac{y_t^2}{2a_t}$\\
   Scale  &  $\frac{\Gamma(\frac{\nu+1}{2})}{\Gamma(\frac{\nu}{2})\sqrt{\pi(\nu-2)e^{a_t}}}\left[1+\frac{y_t^2}{(\nu-2) e^{a_t}}\right]^{-\frac{\nu+1}{2}}$ & $\frac{1}{2}\left[\frac{(\nu+1)y_t^2}{(\nu-2)e^{a_t}+y_t^2}-1  \right]$ & $-\frac{(\nu-2)(\nu+1)y_t^2e^{a_t}}{[(\nu-2)e^{a_t}+y_t^2]^2}$\\
   Count data & $\frac{a_t^{y_t} e^{-a_t}}{y_t!}$ & $\frac{y_t}{a_t}-1$ & $-\frac{y_t}{a_t^2}$\\
   \hline
  \end{tabular}
  \end{adjustbox}
\caption{For each state-space model, we specify the observation density $p(y_t|\alpha_t=a_t)$, the score $\nabla_t(a_t)$ and the Hessian $h_t(a_t)$ functions employed in the filtering recursions of Section \eqref{sec:filtSmooth}.}
\label{tab:filterSpec}
\end{table}

\noindent The variance $q$ is a relevant parameter in this analysis because it is related to the accuracy of the approximation. When $q$ tends to zero, we expect the approximation error being small because the prior density $p(\alpha_t|Y_{t-1})$ tends to a degenerate distribution. On the contrary, when $q$ is large, we expect a deterioration of the approximate filter.\ We consider three scenarios with $q=0.005,0.01,0.05$ 
and $c = 0.001$, $\phi=0.98$. 
The third scenario with $q=0.05$ leads to very erratic latent state dynamics in all the four state-space models.


\afterpage{
\begin{landscape}%
\begin{table}[htb]
\centering
\setlength{\tabcolsep}{3pt}
\renewcommand{\arraystretch}{1}
\begin{tabular}{c|c|ccccc|ccccc|ccccc }
  \hline
   &  & IS &  & Rob.\ recurs.   & QMLE & KF  & IS &  & Rob.\ recurs.    & QMLE & KF & IS & & Rob.\ recurs.    & QMLE & KF\\
 \hline
  & & \multicolumn{5}{c|}{$q=0.005$} & \multicolumn{5}{c|}{$q=0.01$} & \multicolumn{5}{c}{$q=0.05$}\\
 \hline
 \multirow{3}{*}{\shortstack{Location\\ \footnotesize{(Student-$t$)}} }    
 & Prediction   &  0.0120  &   & 0.0123  & $-$ & 0.0135 & 0.0232 &  & 0.0240  & $-$ & 0.0270 &  0.1140 &  & 0.1193 & $-$ & 0.1289 \\
 & Update   &  0.0066  &   & 0.0068  & $-$ & 0.0088 & 0.0132 &  & 0.0134  & $-$ & 0.0176 &  0.0646 &   & 0.0694 & $-$ & 0.0799\\
 & Smoother   &  0.0054  &   & 0.0054  & $-$ & 0.0059 & 0.0099 &   & 0.0104  & $-$ & 0.0115 &  0.0518 &  & 0.0546 & $-$ & 0.0580\\
     \hline
 \multirow{3}{*}{\shortstack{Scale\\ \footnotesize{(Gaussian)}} }     
 & Prediction & 0.0736  &  & 0.0736 & 0.1019  & $-$ & 0.1171 &  & 0.1189  & 0.1638 & $-$ & 0.3247 &  & 0.3519 & 0.4579 & $-$\\
 & Update & 0.0715  &  & 0.0715 & 0.0987  & $-$ & 0.1115 &  & 0.1134  & 0.1592 & $-$ & 0.2857 &  & 0.3143 & 0.4237 & $-$\\
& Smoother & 0.0537  &  & 0.0538 & 0.0896  & $-$ & 0.0760 & & 0.0770  & 0.1182 & $-$ & 0.1823 &  & 0.2099 & 0.2784
 & $-$\\
     \hline
 \multirow{3}{*}{\shortstack{Scale\\ \footnotesize{(Student-$t$)}} }   
& Prediction &   0.0876 &  & 0.0878 & 0.1428  & $-$ &  0.1366 &  & 0.1379 & 0.1982 & $-$ &  0.3806 &  & 0.3960 &  0.5030 & $-$\\
& Update &   0.0859 &  & 0.0860 & 0.1428  & $-$ &  0.1326 &  & 0.1332 & 0.1892 & $-$ &  0.3442 &  & 0.3621 &  0.4724 & $-$\\
 & Smoother &   0.0682 &  & 0.0682 & 0.1382  & $-$ &  0.0950 &  & 0.0955 & 0.1816 & $-$ &  0.2405 &  & 0.2448 &  0.3516 & $-$\\
 	 \hline
 \multirow{3}{*}{\shortstack{Count data\\ \footnotesize{(Poisson)}} }     
& Prediction &  0.0787  &   & 0.0787  & $-$ & $-$  & 0.1391 &   & 0.1408 &  $-$ & $-$ &  0.4291 &    & 0.4659 &  $-$ & $-$\\ 
& Update &  0.0770  &    & 0.0772  & $-$ & $-$  & 0.1311 &  & 0.1325 &  $-$ & $-$ &  0.3947 &    & 0.4340 &  $-$ & $-$\\		
& Smoother &  0.0498  &  & 0.0501  & $-$ & $-$  & 0.0759 &   & 0.0792 &  $-$ & $-$ &  0.2168 &    & 0.2732 &  $-$ & $-$\\
 \hline
 
 \end{tabular}			
 \caption{We report the MSE of the filtered and smoothed estimates of the four state-space models obtained using importance sampling (IS) and the robust filtering and smoothing recursions of Section \eqref{sec:filtSmooth} based on score and Hessian functions (Rob. recurs.).\ For the location model, we also show the MSE of the Kalman filter (KF), whereas for the two stochastic volatility models we report the MSE of the QMLE method of \cite{RUIZ1994289}.}
\label{tab:MCmse}
\end{table}
 \end{landscape}
}

The noise log-variance parameter in the location model is set as $\lambda=\log(5q)$, restricting the noise-to-signal ratio $e^{\lambda}/q$ to be equal to $5$ in each scenario.
The effect of choosing a different value for the noise-to-signal ratio is discussed below. Finally, we set $\nu=5$ in the two models based on Student-$t$ distribution and adopt a parameterization for $t_{\nu}$ guaranteeing that the variance coincides with the scale factor. Table \eqref{tab:filterSpec} reports, for each state-space model, the expression of the observation density $p(y_t|\alpha_t=a_t)$, the score and Hessian functions. 

The optimal filtered and smoothed estimates of $\alpha_t$ are computed using Importance Sampling (IS), as described e.g., in \cite{RICHARD20071385}. To set the importance density, we use the ``Numerically Accelerated Importance Sampling'' (NAIS) method of \cite{NAIS}, implemented by sampling $N=400$ antithetic paths of $\alpha_t$ from the importance density. The IS filtered and smoothed estimates are computed after estimating the static parameters of the four state-space models using the same NAIS algorithm.\ To run the robust recursions of Section \eqref{sec:filtSmooth}, we use the model parameters estimated using the approximate conditional likelihood $\sum_{t=1}^n\log p(y_t|\alpha_t=a_t)$; see Proposition \eqref{prop:pyt}.\  As further benchmarks, we examine the performance of the standard Kalman filter in the location model, and the QMLE method of \cite{RUIZ1994289} in the two stochastic volatility models. 
 
The Monte Carlo study is based on 1000 replications of $n=4000$ observations of the state-space models described above. Each sample is divided in two sub-samples of equal size. The first sub-sample is used to estimate the model parameters by maximum likelihood, whereas the second is used to compute the mean-square-error (MSE) of the filtered (prediction and update) and smoothed estimates of $\alpha_t$.

The results are reported in Table \eqref{tab:MCmse}, which shows the average MSE provided by each method in the three scenarios $q=0.005,0.01,0.05$.
We first note that, compared to the IS method, the average loss of our local approximation is negligible in the scenario $q=0.005$, small (close to 2\%) in the scenario $q=0.01$, and significant (larger than 8\%) in the scenario $q=0.05$. 
We also note that the relative loss of the Kalman filter (in the location model) and QMLE (in stochastic volatility models) is significantly large compared to our methodology. The relative performance of the Kalman filter depends on the non-normality of the observation density, and therefore on the degrees of freedom parameter and the noise-to-signal ratio, which in this analysis are set as $\nu=5$ and $e^{\lambda}/q=5$, respectively. Lower values of $\nu$ or higher values of $e^{\lambda}/q$ would further increase the loss. In contrast, the loss would decrease by increasing $\nu$ or decreasing the ratio $e^{\lambda}/q$. The QMLE exhibits a similar behavior because it applies the standard Kalman filter to a non-Gaussian linearized density.\ Note, indeed, that its relative loss increases in the stochastic volatility model with Student-$t$ distribution.
Not surprisingly, the computational times of our approximate methodology are much lower compared to IS. The ratio between the time required to estimate the static parameters and compute the filtered and smoothed estimates in the two methods ranges from $150$ to $800$.



\begin{table}[htb]
\centering
\setlength{\tabcolsep}{4pt}
\renewcommand{\arraystretch}{1}
\begin{tabular}{c|cccccccc}
  \hline
    & \multicolumn{2}{c}{\shortstack{Location\\ \footnotesize{(Student-$t$)}}} & \multicolumn{2}{c}{\shortstack{Scale\\ \footnotesize{(Gaussian)}}} & \multicolumn{2}{c}{\shortstack{Scale\\ \footnotesize{(Student-$t$)}}}  & \multicolumn{2}{c}{\shortstack{Count data\\ \footnotesize{(Poisson)}}} \\
   \hline
   & \small{Filt} & \small{Par + Filt} & \small{Filt} & \small{Par + Filt}  & \small{Filt} & \small{Par + Filt} & \small{Filt} & \small{Par + Filt}   \\
   \hline
&  \multicolumn{8}{c}{$q=0.005$} \\ 
 \hline 
                     Prediction     & 0.9446  & 0.9489 & 0.9205  & 0.9469  & 0.9264  & 0.9402 &  0.9295  & 0.9332\\
                     Update         & 0.9447  & 0.9490 & 0.9194  & 0.9472  & 0.9251  & 0.9398 &  0.9294 & 0.9331\\
                     Smoother       & 0.9466  & 0.9501 & 0.9171  & 0.9531  & 0.9212  & 0.9422 & 0.9290  & 0.9358\\
            
\hline
 &  \multicolumn{8}{c}{$q=0.01$} \\
 \hline
                     Prediction     & 0.9435  & 0.9483 & 0.9270 & 0.9472  &  0.9291  & 0.9424  & 0.9202 & 0.9232\\
                     Update         & 0.9440  & 0.9486 & 0.9266 & 0.9474  &  0.9285  & 0.9424  & 0.9203 & 0.9238\\
                     Smoother       & 0.9464  & 0.9502 & 0.9187 & 0.9516  &  0.9252  & 0.9476  & 0.9104 & 0.9292\\
                        
                        \hline
&  \multicolumn{8}{c}{$q=0.05$} \\ 
 \hline
                    Prediction     & 0.9181  & 0.9231 &  0.8961 & 0.9415  & 0.9118  & 0.9428 & 0.8450  & 0.8510\\
                    Update         & 0.9319  & 0.9359 &  0.8928 & 0.9418  & 0.9098  & 0.9433 & 0.8442  & 0.8517\\
                    Smoother       & 0.9434  & 0.9450 &  0.8718 & 0.9421  & 0.8982  & 0.9469 & 0.8451  & 0.8540\\
                        \hline
 \end{tabular}			
 
 \caption{We show the average coverage rates of the out-of-sample confidence bands computed around the filtered and smoothed estimates at $95\%$ confidence level for $q=0.005, 0.01, 0.05$. For each state-space model, we report the coverage rates computed by only accounting for filtering uncertainty (Filt) and the ones accounting for both parameter and filtering uncertainty (Par + Filt).}
\label{tab:MCconfBands}
\end{table}

Using the same Monte Carlo sample, we compute approximate out-of-sample confidence bands around the filtered and smoothed estimates using Equation \eqref{eq:filtParUnc}. As discussed in Section \eqref{sub:filtIUnc}, Equation \eqref{eq:filtParUnc} decomposes the conditional variance of the state variables in the sum of two terms representing the effect of filtering and parameter uncertainty. These two terms are computed by simulations, sampling the static parameters from the asymptotic distribution of the maximum-likelihood estimates. We assume an asymptotic  normal distribution with mean equal to the maximum likelihood estimates and variance given by the negative inverse Hessian matrix of the log-likelihood. 
The confidence bands are computed at confidence level $\alpha=0.95$. 
Table \eqref{tab:MCconfBands} reports the average coverage rates of the confidence bands. 
For each state-space model, we show the results obtained by only accounting for filtering uncertainty, i.e., by only estimating the term $\mathbb{E}_{\bs{\theta}}[\mb{P}_t^{\bs{\theta}}]$ in Equation \eqref{eq:filtParUnc}, and those obtained accounting for both parameter and filtering uncertainty, i.e., by estimating the sum $\mathbb{E}_{\bs{\theta}}[\mb{P}_t^{\bs{\theta}}] + \mathbb{E}_{\bs{\theta}}[(\mb{a}_t^{\bs{\theta}}-\mb{a}_t^{\bs{\hat{\theta}}})(\mb{a}_t^{\bs{\theta}}-\mb{a}_t^{\bs{\hat{\theta}}})']$ in Equation \eqref{eq:filtParUnc}.  

For $q=0.005, 0.01$, the proposed methodology provides a very close match to the nominal confidence level when both parameter and filtering uncertainty are taken into account. Neglecting parameter uncertainty underestimates the total uncertainty and leads to a number of exceedances larger than expected. The impact of parameter uncertainty is more accentuated in the two stochastic volatility models, whereas it is less evident in the location model, where the coverage rates do not change substantially when ignoring parameter uncertainty. This is due to the lower variance of the maximum likelihood estimates of the location model. 
 As $q$ increases, the confidence bands computed with our approach tend to become narrow, with coverage rates significantly below the nominal confidence level. This is particularly evident in the count data model, where the coverage rates are substantially lower than 0.95 for $q=0.05$.   
As in the previous analysis, this effect is imputable to the deterioration of the local approximation when the state variance increases. 


\section{Empirical application}
\label{sec:empirics}  

In this section, we compare in an empirical framework the filtered and smoothed estimates of the proposed methodology in order to assess the inferential improvement arising when accounting for contemporaneous and subsequent information in the estimation of the latent states.\
The main problem we face in this analysis is the search for a proxy of the latent state dynamics that could be used to assess the goodness of the filtered and smoothed estimates.\ The use of high-frequency financial returns represents an ideal empirical framework for this purpose because they provide realized measures of volatility that approximate the true unobservable volatility dynamics; see
\cite{AndersenBollerslev}, \cite{AndersenBollerslevDieboldLabys}, \cite{BNSmulti}.\ The idea is to compare the (model free) realized volatility computed using high-frequency data with the (model based) volatility extracted using a time-series model of daily log-returns. The use of a robust (in the sense of \citealt{Patton2011246}) loss function, such as MSE and Qlike, implies that the results we obtain by comparing the filtered and smoothed estimates with the realized volatility proxy are asymptotically equivalent to the ones we would obtain using the true unobservable volatility.\ We perform the analysis using both univariate stochastic volatility models and multivariate dynamic correlation models.


The univariate model of daily log-returns we consider has two stochastic volatility components:
\begin{align}
y_t &= e^{\frac{\theta_t}{2}}\epsilon_t\\
\theta_t &= \omega + \mb{Z}\bs{\alpha}_t\\
\bs{\alpha}_{t+1} &= \mb{T}\bs{\alpha}_{t}+\bs{\eta}_t
\end{align} 
where $\epsilon_t\sim t_{\nu}$, $\mb{Z}=(1,1)$, $\mb{T}=\text{diag}(\phi_1,\phi_2)$, $|\phi_1|, |\phi_2|<1$ and $\bs{\eta}_t$ has zero mean and diagonal variance matrix $\mb{Q}\in\mathbb{R}^{2\times 2}$. Stochastic volatility models with two volatility components have been advocated, for example, by \cite{EngleLee}, \cite{AlizadehSassanDiebold}, \cite{AndersenEtAl2006}, \cite{https://doi.org/10.1111/jtsa.12419}. The two volatility components can be interpreted as representing a ``slow" factor describing the long-term dynamics of volatility and a ``fast" factor describing its short-term behavior. The combination of the two factors leads to an accurate description of the most relevant empirical properties of financial volatility, such as clustering and long memory; see also \cite{Harvey_2013} for an extensive discussion on two-components volatility models.

In order to compute realized volatility, we use 5-minute transaction data of the following set of 17 highly capitalized stocks belonging to the Russell 3000 index: XOM, PFE, MSFT, JPM, IBM, GOOG, GE, DIS, CVX, BAC, AAPL, COP, NVDA, JNJ, PG, KO, LLY. With only three exceptions (GOOG, CVX, COP), the dataset covers the period from 01-12-1999 to 27-09-2013 including 3478 business days. In the case of GOOG, CVX, COP, the first day for which the data is available coincides with 19-08-2004, 10-10-2001, 03-09-2002, respectively. The last available day is instead 27-09-2013 for all the 17 assets. For each day $t$, we recover the daily log-return $y_t$, computed as the difference between closing and opening log-prices, and the realized variance $RV_t$, computed as the sum of 5-minutes squared returns. 

The two-component stochastic volatility model described above is estimated on each daily time-series of log-returns. To run the robust recursions of Section \eqref{sec:filtSmooth}, we assume a Student-$t$ distribution for $p(y_t|\theta_t)$, namely we set:
\begin{equation}
\begin{split}
\log p(y_t|\theta_t) &= \log\Gamma{\left(\frac{\nu+1}{2}\right)}-\log\Gamma\left(\frac{\nu}{2}\right)-\frac{1}{2}\log \pi -\frac{1}{2}\log(\nu-2)\\
&-\frac{{\theta_t}}{2}-\frac{\nu+1}{2}\log\left[1+\frac{y_t^2}{(\nu-2)e^{{\theta}_t}}\right]
\end{split}\medskip
\label{eq:empLogL}
\end{equation}
where ${\theta}_t=\omega+\mb{Z}\bs{\alpha}_t$.\ The score and the Hessian matrix of $\log p(y_t|\theta_t)$ in the filtering and smoothing recursions of Section \eqref{sec:filtSmooth} are thus given by:\bigskip
\begin{align*}
\bs{\nabla}_t(\mb{a}_t)&= \frac{\partial\log p(y_t|\theta_t)}{\partial\theta_t}\bigg\vert_{\theta_t=\omega+\mb{Z}\mb{a}_t}\times\mb{Z}'=\frac{1}{2}\left[\frac{(\nu+1)y_t^2}{(\nu-2)e^{
\theta_t}+y_t^2}-1  \right]\bigg\vert_{\theta_t=\omega+\mb{Z}\mb{a}_t}\times\mb{Z}'\\
\bs{H}_t(\mb{a}_t) &= \frac{\partial^2\log p(y_t|\theta_t)}{\partial\theta_t^2}\bigg\vert_{\theta_t=\omega+\mb{Z}\mb{a}_t}\times\mb{Z}'\mb{Z}=-\frac{(\nu-2)(\nu+1)y_t^2e^{\theta_t}}{[(\nu-2)e^{\theta_t}+y_t^2]^2}\bigg\vert_{\theta_t=\omega+\mb{Z}\mb{a}_t}\times\mb{Z}'\mb{Z}
\end{align*}\smallskip

\begin{figure}
\centering
    \includegraphics[width=0.8\linewidth]{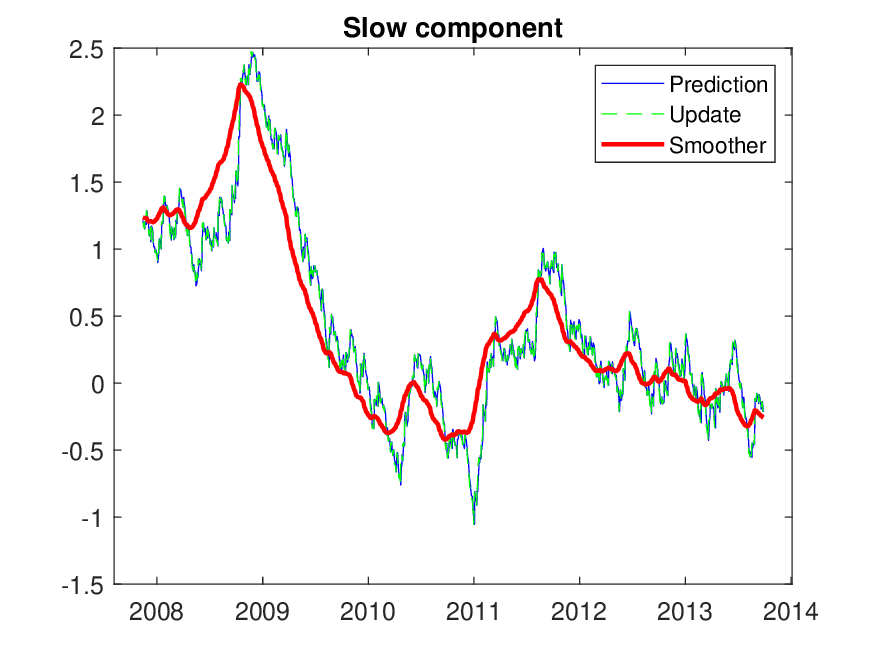}
    \caption{Out-of-sample filtered and smoothed estimates of the slow volatility
component $\bs{\alpha}_t^{(1)}$ of XOM.}
    \label{fig:slowComp}
\end{figure}

\begin{figure}
\centering
    \includegraphics[width=0.8\linewidth]{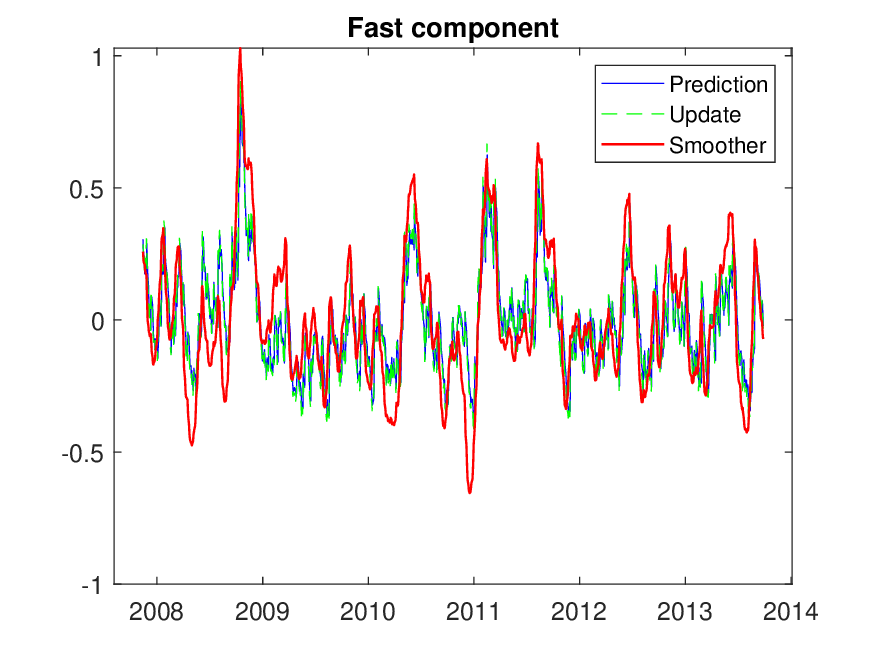}
    \caption{Out-of-sample filtered and smoothed estimates of the fast volatility
component $\bs{\alpha}_t^{(2)}$ of XOM.}
    \label{fig:fastComp}
\end{figure}

\begin{figure}
\centering
    \includegraphics[width=0.8\linewidth]{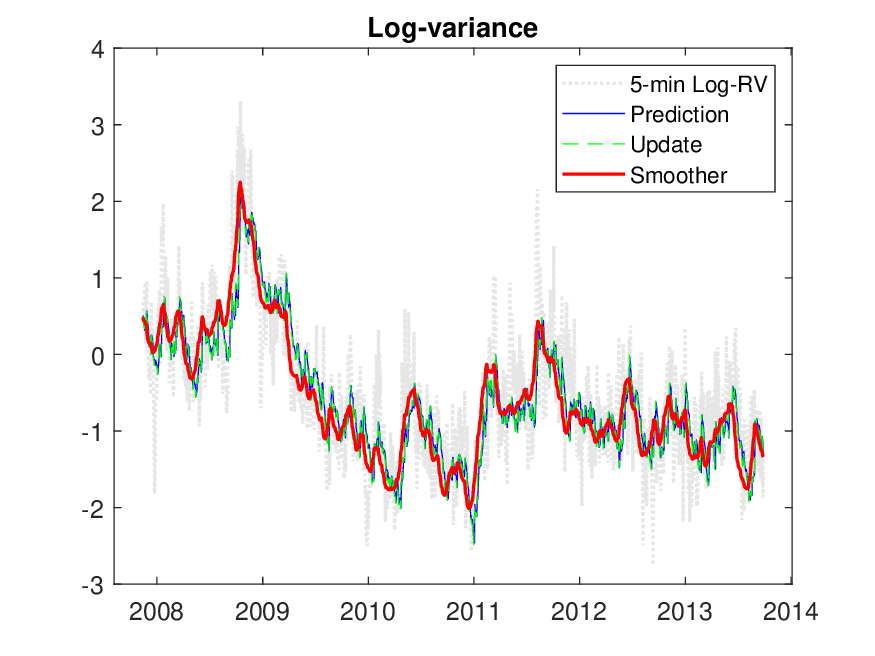}
    \caption{Out-of-sample filtered and smoothed estimates of the logarithmic variance $\theta_t=\omega+\mb{Z}\bs{\alpha}_t$ and logarithm of 5-minute realized variance of XOM.}
    \label{fig:logVar}
\end{figure}

The parameters $\nu$, ${\omega}$, $\mb{T}$, $\mb{Q}$ are estimated in the first sub-sample of $n=2000$ business days by maximizing the approximate log-likelihood $\sum_{t=1}^{2000}\log p(y_t|\theta_t=\omega+\mb{Z}\mb{a}_t)$. The out-of-sample filtered and smoothed estimates are then computed in the sub-sample including the remaining business days.\ In the following, we report the parameter estimates of the XOM time-series:
\begin{align*}
&\hat{\omega}=-1.0091, \quad \hat{\phi}_1 =0.9986,\quad \hat{\phi}_2=0.9354,\quad  \\
&\bs{\hat{Q}}_{11} = 0.0030,\quad \bs{\hat{Q}}_{22} = 0.0198,\quad \hat{\nu} = 9.7201   
\end{align*}
As commonly observed in two-component models, the slow component has large persistence ($\hat{\phi}_1\approx 1$) and lower variance compared to the fast component ($\bs{\hat{Q}}_{11}\ll\bs{\hat{Q}}_{22}$). The parameter estimates of the other time-series behave similarly.
Figures \eqref{fig:slowComp} and \eqref{fig:fastComp} show the out-of-sample filtered and smoothed estimates $\mb{a}_{t}^{(i)}$, $\mb{a}_{t|t}^{(i)}$, $\mb{a}_{t|n}^{(i)}$, $i=1,2$, of the two volatility components of the XOM time-series; Figure \eqref{fig:logVar} shows instead the out-of-sample filtered and smoothed log-variances $\omega + \mb{Z}\mb{a}_t$, $\omega + \mb{Z}\mb{a}_{t|t}$, $\omega + \mb{Z}\mb{a}_{t|n}$ and the logarithm of the 5-minute realized variance.

\begin{table}[h]
\centering
\setlength{\tabcolsep}{12pt}
\renewcommand{\arraystretch}{1}
\begin{tabular}{c|llll}
 \hline
& \multicolumn{2}{c}{MSE} & \multicolumn{2}{c}{Qlike}  \\ 
& In-sample & Out-of-sample & In-sample & Out-of-sample \\
 \hline
    Prediction      & 0.3105$^{(0)}$   &  0.3558$^{(0)}$  &  0.1783$^{(1)}$   & 0.2612$^{(1)}$   \\
    				& 1.0000           &  1.0000          &  1.0000           & 1.0000\\
    Update          & 0.2758$^{(11)}$  &  0.3132$^{(3)}$  &  0.1501$^{(11)}$  & 0.2092$^{(3)}$   \\
                    & 0.8882           &  0.8803          &  0.8418           & 0.8009  \\
    Smoother        & 0.2710$^{(16)}$  &  0.2878$^{(16)}$ &  0.1454$^{(17)}$  & 0.1820$^{(17)}$   \\
                    & 0.8728           &   0.8089         &  0.8155           & 0.6968 \\
\hline
 \end{tabular}			
  \caption{Absolute and relative average in-sample and out-of-sample MSE and Qlike losses of the filtered and smoothed estimates of the 17 time-series of stock returns. We report in parentheses the number of time-series for which the corresponding estimate is included in the MCS test of \cite{MCS} at 95\% confidence level. }
\label{tab:lossEmp}
\end{table}

To assess the extent to which the filtered and smoothed estimates are close to the latent volatility process, we compute the in-sample and out-of-sample MSE and Qlike for the 17 time-series. When computing the two loss functions, we replace the true logarithmic variance $\theta_t$ with the logarithm of the 5-minute realized variance. The results are reported in Table \eqref{tab:lossEmp}, which shows the absolute and relative in-sample and out-of-sample MSE and Qlike losses computed as an average over the 17 time-series. For each estimate, the table reports the corresponding loss functions and the number of time-series for which the estimate is judged to be significantly better than others according to the Model Confidence Set (MCS) test of \cite{MCS} at $95\%$ confidence level. 
We first note a significant drop of both MSE and Qlike when accounting for contemporaneous information in the estimation of volatility. The drop of the two loss functions goes from  $\sim$11\% (MSE, in-sample) to $\sim$20\% (Qlike, out-of-sample). Accounting also for subsequent information leads to a further drop, especially in the out-of-sample estimates. For example, the Qlike loss of the out-of-sample smoother is $\sim$30\% smaller than the Qlike loss of the predictive filter.\ With only one exception, the latter is never included in the model confidence set.\ 

In light of such results, we expect filtering uncertainty being non-degenerate and having a significant effect on the computation of confidence bands. To verify this, we compute the in-sample and out-of-sample confidence bands around the filtered and smoothed estimates by only accounting for parameter uncertainty, i.e., by only estimating the term $ \mathbb{E}_{\bs{\theta}}[(\mb{a}_t^{\bs{\theta}}-\mb{a}_t^{\bs{\hat{\theta}}})(\mb{a}_t^{\bs{\theta}}-\mb{a}_t^{\bs{\hat{\theta}}})']$ in Equation \eqref{eq:filtParUnc}, and then accounting for both parameter and filtering uncertainty, i.e., by estimating the sum $\mathbb{E}_{\bs{\theta}}[\mb{P}_t^{\bs{\theta}}] + \mathbb{E}_{\bs{\theta}}[(\mb{a}_t^{\bs{\theta}}-\mb{a}_t^{\bs{\hat{\theta}}})(\mb{a}_t^{\bs{\theta}}-\mb{a}_t^{\bs{\hat{\theta}}})']$ in Equation \eqref{eq:filtParUnc}.\ Table \eqref{tab:EmpConfBands} shows the average coverage rates of the 95\% confidence bands. We note that neglecting filtering uncertainty and taking into account only the uncertainty of the maximum likelihood estimates leads to very narrow confidence bands with a number of exceedances significantly larger than what expected based on the choice of the nominal confidence level.  
On the contrary, including the effect of filtering uncertainty leads to a very close match to the nominal confidence level. In principle, we could also compute the confidence bands by only accounting for filtering uncertainty, as in the Monte Carlo analysis of Section \eqref{sec:MonteCarlo}. This would also lead to narrow confidence bands underestimating the total uncertainty. We conclude that both parameter and filtering uncertainty are relevant and must be taken into account when computing confidence bands in state-space models.

\begin{table}[h]
\centering
\setlength{\tabcolsep}{12pt}
\renewcommand{\arraystretch}{1}
\begin{tabular}{c|cccc}
 \hline
 & \multicolumn{2}{c}{Parameter uncertainty} & \multicolumn{2}{c}{Parameter + filtering uncertainty} \\ 
 & In-sample & Out-of-sample & In-sample & Out-of-sample \\
 \hline
   Prediction  &    0.7005  &  0.7016   &  0.9489  &  0.9392    \\
   Update      &    0.7283  &  0.7292   &  0.9572  &  0.9470\\
   Smoother    &    0.7501  &  0.7570   &  0.9407  &  0.9479 \\                    
\hline
 \end{tabular}			
  \caption{We show the average coverage rates of the in-sample and out-of-sample confidence bands around the filtered and smoothed estimates at $95\%$ confidence level. The coverage rates are computed as an average over the 17 time-series of stock returns. We report the coverage rates computed by only accounting for parameter uncertainty and the ones accounting for both parameter and filtering uncertainty.}
\label{tab:EmpConfBands}
\end{table}

We now perform the same exercise using a multivariate model of dynamic correlations.\ We assume that the vector of log-returns $\mb{y}_t\in\mathbb{R}^p$ is generated by a Student-$t$ distribution with $\nu$ degrees of freedom and a time-varying covariance matrix $\boldsymbol{\Sigma}(\boldsymbol{\alpha}_t)\in\mathbb{R}^{p\times p}$ depending on a stochastic vector $\boldsymbol{\alpha}_t\in\mathbb{R}^m$:  
\begin{align}
\mb{y}_t|\boldsymbol{\alpha}_t &\sim \frac{\Gamma\left(\frac{\nu+p}{2}\right)}{\Gamma\left(\frac{\nu}{2}\right)[(\nu-2)\pi]^{p/2}\text{det}[\boldsymbol{\Sigma}(\boldsymbol{\alpha}_t)]^{1/2}}\left(1+ \frac{\mb{y}_t'\boldsymbol{\Sigma}(\boldsymbol{\alpha}_t)\mb{y}_t}{\nu-2}\right)^{-\frac{\nu+p}{2}}\label{eq:StudMult}\\
\bs{\alpha}_{t+1} &= \mb{T}\bs{\alpha}_{t}+\bs{\eta}_t
\end{align}
where $\mb{T}\in\mathbb{R}^{m\times m}$ is diagonal and $\bs{\eta}_t$ has zero mean and diagonal covariance matrix $\mb{Q}\in\mathbb{R}^{m\times m}$.\ We use the following decomposition of the covariance matrix:
\begin{equation*}
\boldsymbol{\Sigma}(\boldsymbol{\alpha}_t) = \mb{D}(\boldsymbol{\alpha}_t)\mb{R}(\boldsymbol{\alpha}_t)\mb{D}(\boldsymbol{\alpha}_t)
\end{equation*}
where $\mb{D}(\boldsymbol{\alpha}_t)$ is a diagonal matrix containing the time-varying conditional standard deviations of $\mb{y}_t$ and $\mb{R}(\boldsymbol{\alpha}_t)$ is a correlation matrix.\ In order to ensure the positive-definiteness of $\boldsymbol{\Sigma}(\boldsymbol{\alpha}_t)$, we write $\mb{D}_{ii}(\boldsymbol{\alpha}_t)=e^{\bs{\alpha}_{t,i}}$, for $i=1,\dots,p$, and $\mb{R}(\boldsymbol{\alpha}_t)=\mb{Z}(\bs{\alpha}_t)'\mb{Z}(\bs{\alpha}_t)$, where $\mb{Z}(\bs{\alpha}_t)$ is an upper-triangular matrix of hyperspherical coordinates; see, for example, \cite{GAS2}, where a score-driven model having the same measurement equation is proposed.\ In order to simplify the estimation, the two matrices $\mb{T}$, $\mb{Q}$ are parameterized as follows:\medskip
\begin{equation*}
\mb{T}=
\begin{pmatrix}
\phi_1 \mb{I}_p & \mb{0}\\
\mb{0} & \phi_2 \mb{I}_{\frac{p(p-1)}{2}}
\end{pmatrix},\quad
\mb{Q}=
\begin{pmatrix}
q_1 \mb{I}_p & \mb{0}\\
\mb{0} & q_2 \mb{I}_{\frac{p(p-1)}{2}}
\end{pmatrix}
\medskip
\end{equation*}
where $|\phi_1|,|\phi_2|<1$ and $q_1,q_2>0$. Both $\mb{T}$ and $\mb{Q}$ thus depend on two parameters, one describing the persistence of the volatility processes, and another one describing the persistence of the correlation dynamics. 

To perform the analysis, we use 5-minute transaction data of a set of 150 liquid and illiquid stocks belonging to the Russell 3000 index. The period we consider goes from 27-09-2005 to 27-09-2013. This sample is shorter compared to the one used in the univariate analysis because we have pre-filtered the data in order to select a sub-sample for which the data of all 150 stocks is available.\ We run the analysis using five groups of assets characterized by different levels of liquidity.\ Specifically, among the universe of 150 assets, we select five groups of $p=5$ assets using the following procedure:
\begin{itemize}
\item We compute, for all assets, the average fraction of zero returns $\bar{f_i}$, $i=1,2,\dots,250$, using the 5-minute log-prices.
\item We compute the quartile separators $Q_1$, $Q_2$, $Q_j$ of the sample $[\bar{f}_1,\bar{f}_2,\dots,\bar{f}_{250}]$.
\item Four groups of assets are obtained by  randomly picking $p=5$ stocks with, respectively, $\bar{f}_i< Q_1$, $Q_1\le \bar{f}_i< Q_2$, $Q_2\le \bar{f}_i< Q_3$,$ \bar{f}_i\ge Q_3$. We call such groups \textit{very high, high, medium, low} liquidity. A fifth group is selected by randomly picking $p=5$ assets among the entire universe of 150 assets. We call this group \textit{mixed} liquidity.
\end{itemize}

The dynamic conditional volatility model in Equation \eqref{eq:StudMult} is estimated for the five groups of assets using daily log-returns and the filtered and smoothed estimates of $\bs{\Sigma}(\bs{\Theta}_t)$ are compared with the realized covariance estimator of \cite{BNSmulti} computed using 5-minute returns. The results are reported in Table \eqref{tab:multiEmp}, which shows the MSE and Qlike losses of the five portfolios.\ As in the univariate exercise above, we observe a significant drop of the two loss functions when contemporaneous and/or subsequent information is accounted for in the estimation of $\bs{\Sigma}(\bs{\Theta}_t)$.\ The drop is observed for all the five groups of assets and it is more pronounced when looking at the MSE loss. For example, in the case of the ``high liquidity" portfolio, the MSE diminishes by $\sim 6\%$ when accounting for contemporaneous information, and by $\sim 16\%$ when subsequent information is considered.\ Note also that, for each group and for both loss functions, the MCS test selects only the smoothed estimates as the closest to the true realized covariance.\ Such results further highlight the informational content of contemporaneous and subsequent information and its impact on the estimation of volatility and correlations.

\begin{table}[h]
\centering
\setlength{\tabcolsep}{8pt}
\renewcommand{\arraystretch}{1}
\begin{tabular}{c|lllll}
 \hline
Liquidity & \textit{Very high} & \textit{High} & \textit{Medium} & \textit{Low} & \textit{Mixed} \\ 
\hline
 & \multicolumn{5}{c}{MSE}  \\
 \hline
    Prediction      & 2.463$\times 10^{-6}$   &  2.842$\times 10^{-7}$  &  6.978$\times 10^{-7}$   & 8.002$\times 10^{-7}$  & 1.015$\times 10^{-7}$   \\
    				& 1.000           &  1.000          &  1.000           & 1.000  & 1.000\\
    Update          & 2.369$\times 10^{-6}$  &  2.695$\times 10^{-7}$  &  6.595$\times 10^{-7}$  & 7.867$\times 10^{-7}$ &  9.997$\times 10^{-8}$   \\
                    & 0.962           &  0.948          &   0.945           & 0.983 & 0.984  \\
    Smoother        & 2.281$\times 10^{-6\star}$  &  2.389$\times 10^{-7\star}$ &  6.214$\times 10^{-7\star}$  & 7.352$\times 10^{-7\star}$ & 9.537$\times 10^{-8\star}$    \\
                    & 0.926           &   0.840         &  0.890           & 0.918 & 0.938 \\
\hline
& \multicolumn{5}{c}{Qlike}  \\
 \hline
    Prediction      & $-$31.896   &  $-$37.008  &  $-$34.095   & $-$34.034  &  $-$37.375 \\
    				& 1.000           &  1.000          &  1.000           & 1.000 & 1.000\\
    Update          & $-$32.190  &  $-$37.208  &  $-$34.309  & $-$34.190  & $-$37.678  \\
                    & 0.990          &  0.994          &  0.993           & 0.995  & 0.992  \\
    Smoother        & $-$32.334$^{\star}$  &  $-$37.331$^{\star}$ &  $-$34.435$^{\star}$  & $-$34.338$^{\star}$  & $-$37.770   \\
                    & 0.986           &   0.991         &  0.990       & 0.991  & 0.989 \\
\hline
 \end{tabular}			
  \caption{Absolute and relative average in-sample and out-of-sample MSE and Qlike losses of the filtered and smoothed estimates of the matrix $\bs{\Sigma}(\bs{\Theta}_t)$ for the five groups of assets characterized by different liquidity levels. The star indicates that the corresponding estimate is included in the MCS test of \cite{MCS} at 95\% confidence level. }
\label{tab:multiEmp}
\end{table}

\section{Conclusions}
\label{sec:Conclusions}

In conclusion, this paper presents a general approximate filtering and smoothing methodology for nonlinear and non-Gaussian state-space models.\ The main property of the methodology is the updating mechanism of the mean and variance estimates based on the score and Hessian matrix of the observation density. When the state-space model deviates from the linear and Gaussian specification, this updating mechanism provides robust state estimates by winsorizing the extremely large observations generated by the non-Gaussian density.\ The methodology also allows a straightforward computation of in-sample and out-of-sample confidence bands around the state estimates reflecting the combination of filtering and parameter uncertainty.

Our method generalizes other nonlinear filtering techniques based on the score and the Hessian matrix of the observation density along different directions.\ For example, using a perturbation approach, we show that the filtering recursions can be derived by relaxing the assumption of a Gaussian prior density, which is at the basis of previous derivations of this class of filters. Since in nonlinear and non-Gaussian state-space models the prior density is generally non-Gaussian, our results show that the methodology can be applied in realistic conditions where the prior density deviates from normality. This is confirmed by the Monte Carlo analysis, where the loss incurred by our approximate technique with respect to exact simulation-based methods is found to be small in a wide range of scenarios. Moreover, we assume, as a data generating process, a general specification including location, scale and other types of state-space models. This represents a relevant progress with respect to the extant literature, where filters based on the score and Hessian matrix are derived assuming a location model as a data generating process.

Another key contribution of this work is the construction of in-sample and out-of-sample confidence bands that encapsulate both parameter and filtering uncertainty, providing a more comprehensive understanding of the model's reliability.\ Both kinds of uncertainties are shown to be relevant using both simulated and empirical data.

\newpage


\bibliography{biblio}
\bibliographystyle{elsarticle-harv}

\clearpage

\clearpage


\appendix
\appendixpageoff
\numberwithin{equation}{section}

\huge{\textbf{Appendix}}
\normalsize
\vspace{0.5cm}

%

\noindent In the following, in order to simplify the notation, we use $\nabla_t$ in place of $\nabla(a_t)$ and $h_t$ in place of $h_t(a_t)$.

\section{Proof of Proposition \eqref{prop:pyt}}
\label{app:prop:pyt}
We can write the conditional density $p(y_t|Y_{t-1})$ as:
\begin{equation}
p(y_t|Y_{t-1})=\int_{-\infty}^{+\infty} p(y_t|\alpha_t)p(\alpha_t|Y_{t-1})d\alpha_t
\label{eq:app:pyt}
\end{equation}
Let $\delta\ge \bar{\delta}$, where $\bar{\delta}$ is defined in Assumption \eqref{ass:2}. Expanding $p(y_t|\alpha_t)$ at first order in a neighborhood of center $a_t$ and radius $\delta$, we obtain:
\begin{equation}
p(y_t|\alpha_t) = p(y_t|\alpha_t)|_{a_t} + \frac{\partial p(y_t|\alpha_t)}{\partial\alpha_t}\bigg |_{a_t}(\alpha_t-a_t)+ g(\alpha_t)
\label{eq:app:taylor1st}
\end{equation} 
where $g(\alpha_t)$ is a function such that $|g(\alpha_t)|\le M(\delta)(\alpha_t-a_t)^2$ for each $\alpha_t$ belonging to the neighborhood. Because of Assumption \eqref{ass:1}, the constant $M(\delta)$ is given by $M(\delta)=\frac{1}{2}\sup_{|\alpha_t-a_t|\le \delta}\left |\frac{\partial^2 p(y_t|\alpha_t)}{\partial\alpha_t^2}\right |$. We can thus write the integral in Equation \eqref{eq:app:pyt} as:
\begin{align*}
p(y_t|Y_{t-1})=& \int_{-\infty}^{+\infty}\left[p(y_t|\alpha_t)|_{a_t} + \frac{\partial p(y_t|\alpha_t)}{\partial\alpha_t}\bigg |_{a_t}(\alpha_t-a_t)+ g(\alpha_t)\right]p(\alpha_t|Y_{t-1})d\alpha_t\\
=\ & p(y_t|\alpha_t)|_{a_t} + \int_{-\infty}^{+\infty} g(\alpha_t)p(\alpha_t|Y_{t-1})d\alpha_t
\end{align*}
Let us decompose the integral over $g(\alpha_t)$ as:
\begin{align*}
\int_{-\infty}^{+\infty} g(\alpha_t)p(\alpha_t|Y_{t-1})d\alpha_t & = \int_{-\infty}^{a_t-\delta} g(\alpha_t)p(\alpha_t|Y_{t-1})d\alpha_t \\  + & \int_{a_t-\delta}^{a_t+\delta} g(\alpha_t)p(\alpha_t|Y_{t-1})d\alpha_t +\int_{a_t+\delta}^{+\infty} g(\alpha_t)p(\alpha_t|Y_{t-1})d\alpha_t
\end{align*}
Observe that the second integral can be bounded as follows:
\begin{align*}
\left | \int_{a_t-\delta}^{a_t+\delta} g(\alpha_t)p(\alpha_t|Y_{t-1})d\alpha_t \right| & \le \int_{a_t-\delta}^{a_t+\delta} \left | g(\alpha_t)\right |p(\alpha_t|Y_{t-1})d\alpha_t \\
& \le \int_{a_t-\delta}^{a_t+\delta} M(\delta)(\alpha_t-a_t)^2p(\alpha_t|Y_{t-1})d\alpha_t & \\ 
& \le M(\delta)p_t
\end{align*}
To bound the first and third integrals, we use the condition in Assumption \eqref{ass:2}. Let us focus on the third integral:
\begin{align*}
\left | \int_{a_t+\delta}^{+\infty} g(\alpha_t)p(\alpha_t|Y_{t-1})d\alpha_t\right | & \le \int_{a_t+\delta}^{+\infty} \left |g(\alpha_t)\right |p(\alpha_t|Y_{t-1})d\alpha_t  \\
& = \int_{a_t+\delta}^{+\infty} \left| p(y_t|\alpha_t)-p(y_t|\alpha_t)|_{a_t}-\frac{\partial p(y_t|\alpha_t)}{\partial\alpha_t}\bigg |_{a_t}(\alpha_t-a_t)\right |p(\alpha_t|Y_{t-1})d\alpha_t \\
& \le \int_{a_t+\delta}^{+\infty}\frac{\sup_{\alpha_t\ge a_t+\delta}p(y_t|\alpha_t)+p(y_t|\alpha_t)|_{a_t}}{(\alpha_t-a_t)^{3+\beta}}d\alpha_t \\
&+ \int_{a_t+\delta}^{+\infty} \left |\frac{\partial p(y_t|\alpha_t)}{\partial\alpha_t}\bigg |_{a_t}\right |\frac{1}{(\alpha_t-a_t)^{2+\beta}}d\alpha_t
\end{align*}
where the second inequality is due to the boundedness of $p(y_t|\alpha_t)$ and to the restriction $\delta\ge \bar{\delta}$. Computing the two integrals we get:
\begin{align*}
\left | \int_{a_t+\delta}^{+\infty} g(\alpha_t)p(\alpha_t|Y_{t-1})d\alpha_t\right | & \le \frac{1}{2+\beta}\frac{\sup_{\alpha_t\ge a_t+\delta}p(y_t|\alpha_t)+p(y_t|\alpha_t)|_{a_t}}{\delta^{2+\beta}} \\
& +  \frac{1}{1+\beta}\left |\frac{\partial p(y_t|\alpha_t)}{\partial\alpha_t}\bigg |_{a_t}\right |\frac{1}{\delta^{1+\beta}}
\end{align*}
Similar computations lead to:
\begin{align*}
\left | \int_{-\infty}^{a_t-\delta} g(\alpha_t)p(\alpha_t|Y_{t-1})d\alpha_t\right | & \le \frac{1}{2+\beta}\frac{\sup_{\alpha_t\le a_t-\delta}p(y_t|\alpha_t)+p(y_t|\alpha_t)|_{a_t}}{\delta^{2+\beta}} \\
& +  \frac{1}{1+\beta}\left |\frac{\partial p(y_t|\alpha_t)}{\partial\alpha_t}\bigg |_{a_t}\right |\frac{1}{\delta^{1+\beta}}
\end{align*}
We choose $\tilde{\delta}$ as the lowest $\delta\ge \bar{\delta}$ for which the following inequality holds:
\begin{equation*}
\frac{1}{2+\beta}\frac{\sup_{\alpha_t\le a_t-\delta}p(y_t|\alpha_t)+\sup_{\alpha_t\ge a_t+\delta}p(y_t|\alpha_t)+2p(y_t|\alpha_t)|_{a_t}}{\delta^{2+\beta}} + \frac{2}{1+\beta}\left |\frac{\partial p(y_t|\alpha_t)}{\partial\alpha_t}\bigg |_{a_t}\right |\frac{1}{\delta^{1+\beta}}\le\gamma
\end{equation*}
Setting $\xi_t=\int_{-\infty}^{+\infty} g(\alpha_t)p(\alpha_t|Y_{t-1})d\alpha_t$, we can therefore write $p(y_t|Y_{t-1})$ as:
\begin{equation*}
p(y_t|Y_{t-1}) = p(y_t|\alpha_t)|_{a_t}+ \xi_t
\end{equation*}
where $|\xi_t|\le \frac{1}{2}\sup_{|\alpha_t-a_t|\le \tilde{\delta}}\left |\frac{\partial^2 p(y_t|\alpha_t)}{\partial\alpha_t^2}\right |p_t+\gamma$. \begin{flushright}
$\square$
\end{flushright}

\section{Proof of Theorem \eqref{prop:att}}
\label{app:prop:att}

The first moment of $p(\alpha_t|Y_t)$ can be computed as:
\begin{equation*}
a_{t|t} = \int_{-\infty}^{+\infty}\alpha_t p(\alpha_t|Y_t)d\alpha_t=\int_{-\infty}^{+\infty}\alpha_t\frac{p(y_t|\alpha_t)p(\alpha_t|Y_{t-1})}{p(y_t|Y_{t-1})}d\alpha_t
\label{eq:app:att}
\end{equation*}
As in Appendix \eqref{app:prop:pyt}, let us consider the first order expansion of $p(y_t|\alpha_t)$ in Equation \eqref{eq:app:taylor1st}:
\begin{equation*}
p(y_t|\alpha_t) = p(y_t|\alpha_t)|_{a_t} + \frac{\partial p(y_t|\alpha_t)}{\partial\alpha_t}\bigg |_{a_t}(\alpha_t-a_t)+ g(\alpha_t)
\end{equation*}
where $|g(\alpha_t)|\le M(\delta)(\alpha_t-a_t)^2$ for each $\alpha_t$ belonging to a neighborhood of center $a_t$ and radius $\delta$. The constant $M(\delta)$ is given by $M(\delta)=\frac{1}{2}\sup_{|\alpha_t-a_t|\le \delta}\left |\frac{\partial^2 p(y_t|\alpha_t)}{\partial\alpha_t^2}\right |$ thanks to Assumption \eqref{ass:1}. We set $\delta\ge\bar{\delta}$, where $\bar{\delta}$ is defined in Assumption \eqref{ass:3}.
We can thus write:
\begin{equation*}
a_{t|t} = \frac{p(y_t| \alpha_t)|_{a_t}}{p(y_t|Y_{t-1})}\int_{-\infty}^{+\infty}\alpha_t\left[1 + \nabla_t(\alpha_t-a_t)+ \frac{g(\alpha_t)}{p(y_t|\alpha_t)|_{a_t}} \right]p(\alpha_t|Y_{t-1})d\alpha_t
\end{equation*}
Let us compute the integral of the first two terms:
\begin{align*}
&\quad \int_{-\infty}^{+\infty}\alpha_t\left[1 + \nabla_t(\alpha_t-a_t)\right]p(\alpha_t|Y_{t-1})d\alpha_t  \\
& = \int_{-\infty}^{+\infty}(a_t +\alpha_t-a_t)\left[1 + \nabla_t(\alpha_t-a_t)\right]p(\alpha_t|Y_{t-1})d\alpha_t\\
& =  \int_{-\infty}^{+\infty}\left[a_t + a_t\nabla_t(\alpha_t-a_t) + (\alpha_t-a_t) + \nabla_t(\alpha_t-a_t)^2\right]p(\alpha_t|Y_{t-1})d\alpha_t \\
& = a_t + \nabla_t p_t
\end{align*}
where the terms in $(\alpha_t-a_t)$ vanish when integrating with respect to $p(\alpha_t|Y_{t-1})d\alpha_t$. \\
We now consider the integral of the last term. Observe that we can write:
\begin{align}
&\quad \int_{-\infty}^{+\infty}\alpha_t\frac{g(\alpha_t)}{p(y_t|\alpha_t)|_{a_t}}p(\alpha_t|Y_{t-1})d\alpha_t\\
& = \frac{1}{p(y_t|\alpha_t)|_{a_t}}\int_{-\infty}^{+\infty}(\alpha_t-a_t+a_t)g(\alpha_t)p(\alpha_t|Y_{t-1})d\alpha_t\\
& = \frac{1}{p(y_t|\alpha_t)|_{a_t}}\left[\int_{-\infty}^{+\infty}(\alpha_t-a_t)g(\alpha_t)p(\alpha_t|Y_{t-1})d\alpha_t + \int_{-\infty}^{+\infty}a_tg(\alpha_t)p(\alpha_t|Y_{t-1})d\alpha_t\right]
\label{eq:app:att1}
\end{align}
The first integral in Equation \eqref{eq:app:att1} can be decomposed as:
\begin{align*}
&\int_{-\infty}^{+\infty}(\alpha_t-a_t)g(\alpha_t)p(\alpha_t|Y_{t-1})d\alpha_t=\int_{-\infty}^{a_t-\delta}(\alpha_t-a_t)g(\alpha_t)p(\alpha_t|Y_{t-1})d\alpha_t \\
&+\int_{a_t-\delta}^{a_t+\delta}(\alpha_t-a_t)g(\alpha_t)p(\alpha_t|Y_{t-1})d\alpha_t + \int_{a_t+\delta}^{+\infty}(\alpha_t-a_t)g(\alpha_t)p(\alpha_t|Y_{t-1})d\alpha_t
\end{align*}
The integral over the neighborhood of center $a_t$ and radius $\delta$ can be bounded as follows:
\begin{align*}
\left | \int_{a_t-\delta}^{a_t+\delta}(\alpha_t-a_t)g(\alpha_t)p(\alpha_t|Y_{t-1})d\alpha_t \right|& \le \int_{a_t-\delta}^{a_t+\delta}|\alpha_t-a_t||g(\alpha_t)|p(\alpha_t|Y_{t-1})d\alpha_t \\
& \le M(\delta) \int_{a_t-\delta}^{a_t+\delta}|\alpha_t-a_t)|^3 p(\alpha_t|Y_{t-1})d\alpha_t \\
& \le M(\delta) \beta_t^{(3)}
\end{align*}
Now, let us consider the integral from $a_t+\delta$ to $+\infty$ and observe that we can write it as:
\begin{align*}
&\quad  \int_{a_t+\delta}^{+\infty}(\alpha_t-a_t)g(\alpha_t)p(\alpha_t|Y_{t-1})d\alpha_t \\
=&\int_{a_t+\delta}^{+\infty}(\alpha_t-a_t)\left[p(y_t|\alpha_t)-p(y_t|\alpha_t)|_{a_t}-\frac{\partial p(y_t|\alpha_t)}{\partial\alpha_t}\bigg |_{a_t}(\alpha_t-a_t)\right]   p(\alpha_t|Y_{t-1})d\alpha_t
\end{align*}
We can bound this integral using the condition in Assumption \eqref{ass:3} and the boundedness of $p(y_t|\alpha_t)$ in Assumption \eqref{ass:1}:
\begin{align*}
&\quad \left| \int_{a_t+\delta}^{+\infty}(\alpha_t-a_t)\left[p(y_t|\alpha_t)-p(y_t|\alpha_t)|_{a_t}-\frac{\partial p(y_t|\alpha_t)}{\partial\alpha_t}\bigg |_{a_t}(\alpha_t-a_t)\right]   p(\alpha_t|Y_{t-1})d\alpha_t \right| \\
& \le \int_{a_t+\delta}^{+\infty}\frac{\sup_{\alpha_t\ge a_t+\delta} p(y_t|\alpha_t) + p(y_t|\alpha_t)|_{a_t}}{(\alpha_t-a_t)^{3+\beta}}d\alpha_t + \int_{a_t+\delta}^{+\infty}\frac{\left|\frac{\partial p(y_t|\alpha_t)}{\partial\alpha_t}\bigg |_{a_t}\right|}{(\alpha_t-a_t)^{2+\beta}}d\alpha_t\\
& = \frac{1}{2+\beta}\frac{\sup_{\alpha_t\ge a_t+\delta} p(y_t|\alpha_t) + p(y_t|\alpha_t)|_{a_t}}{\delta^{2+\beta}} + 
\frac{1}{1+\beta}\frac{\left|\frac{\partial p(y_t|\alpha_t)}{\partial\alpha_t}\bigg |_{a_t}\right|}{\delta^{1+\beta}}
\end{align*}
Similarly, the integral from $-\infty$ to $a_t$ can be bounded as follows:
 \begin{align*}
&\quad \left| \int_{a_t+\delta}^{+\infty}(\alpha_t-a_t)\left[p(y_t|\alpha_t)-p(y_t|\alpha_t)|_{a_t}-\frac{\partial p(y_t|\alpha_t)}{\partial\alpha_t}\bigg |_{a_t}(\alpha_t-a_t)\right]   p(\alpha_t|Y_{t-1})d\alpha_t \right| \\
& \le  \frac{1}{2+\beta}\frac{\sup_{\alpha_t\le a_t-\delta} p(y_t|\alpha_t) + p(y_t|\alpha_t)|_{a_t}}{\delta^{2+\beta}} + 
\frac{1}{1+\beta}\frac{\left|\frac{\partial p(y_t|\alpha_t)}{\partial\alpha_t}\bigg |_{a_t}\right|}{\delta^{1+\beta}}
\end{align*}
Thus, we have:
\begin{align*}
&\quad\quad \left| \int_{-\infty}^{+\infty}(\alpha_t-a_t)g(\alpha_t)p(\alpha_t|Y_{t-1})d\alpha_t\right|\\
&\le M(\delta)\beta_t^{(3)} + \frac{1}{2+\beta}\frac{\sup_{\alpha_t\ge a_t+\delta} p(y_t|\alpha_t)+\sup_{\alpha_t\le a_t-\delta} p(y_t|\alpha_t) + 2p(y_t|\alpha_t)|_{a_t}}{\delta^{2+\beta}} + 
\frac{2}{1+\beta}\frac{\left|\frac{\partial p(y_t|\alpha_t)}{\partial\alpha_t}\bigg |_{a_t}\right|}{\delta^{1+\beta}}
\end{align*}
We now need to bound the second integral in Equation \eqref{eq:app:att1}. As before, let us decompose it as:
\begin{align*}
&\quad\int_{-\infty}^{+\infty}a_tg(\alpha_t)p(\alpha_t|Y_{t-1})d\alpha_t =\int_{-\infty}^{a_t-\delta}a_tg(\alpha_t)p(\alpha_t|Y_{t-1})d\alpha_t  \\
&+\int_{a_t-\delta}^{a_t+\delta}a_tg(\alpha_t)p(\alpha_t|Y_{t-1})d\alpha_t + \int_{a_t+\delta}^{+\infty}a_tg(\alpha_t)p(\alpha_t|Y_{t-1})d\alpha_t 
\end{align*}
As in the previous step, these integrals can be bounded using Assumption \eqref{ass:3} and the boundedness of $p(y_t|\alpha_t)$ in Assumption \eqref{ass:1}. Simple computations lead to:
\begin{align*}
&\quad\quad \left| \int_{-\infty}^{+\infty}a_tg(\alpha_t)p(\alpha_t|Y_{t-1})d\alpha_t\right|\\
&\le M(\delta)|a_t|p_t + \frac{|a_t|}{3+\beta}\frac{\sup_{\alpha_t\ge a_t+\delta} p(y_t|\alpha_t)+\sup_{\alpha_t\le a_t-\delta} p(y_t|\alpha_t) + 2p(y_t|\alpha_t)|_{a_t}}{\delta^{3+\beta}} + 
\frac{2|a_t|}{2+\beta}\frac{\left|\frac{\partial p(y_t|\alpha_t)}{\partial\alpha_t}\bigg |_{a_t}\right|}{\delta^{2+\beta}}
\end{align*}
Let us now choose $\tilde{\delta}$ as the lowest $\delta\ge \bar{\delta}$ satisfying the following inequality:
\begin{align*}
&\quad \frac{1}{p(y_t|\alpha_t)|_{a_t}} \left[ \frac{1}{2+\beta}\frac{\sup_{\alpha_t\ge a_t+\delta} p(y_t|\alpha_t)+\sup_{\alpha_t\le a_t-\delta} p(y_t|\alpha_t) + 2p(y_t|\alpha_t)|_{a_t}}{\delta^{2+\beta}} + 
\frac{2}{1+\beta}\frac{\left|\frac{\partial p(y_t|\alpha_t)}{\partial\alpha_t}\bigg |_{a_t}\right|}{\delta^{1+\beta}} + \right. \\
&\left. +  \frac{|a_t|}{3+\beta}\frac{\sup_{\alpha_t\ge a_t+\delta} p(y_t|\alpha_t)+\sup_{\alpha_t\le a_t-\delta} p(y_t|\alpha_t) + 2p(y_t|\alpha_t)|_{a_t}}{\delta^{3+\beta}} + 
\frac{2|a_t|}{2+\beta}\frac{\left|\frac{\partial p(y_t|\alpha_t)}{\partial\alpha_t}\bigg |_{a_t}\right|}{\delta^{2+\beta}}\right]\le \gamma
\end{align*}
Setting $\chi_t=\int_{-\infty}^{+\infty}\alpha_t\frac{g(\alpha_t)}{p(y_t|\alpha_t)|_{a_t}}p(\alpha_t|Y_{t-1})d\alpha_t$, we have that:
\begin{equation*}
a_{t|t}= \frac{p(y_t| \alpha_t)|_{a_t}}{p(y_t|Y_{t-1})}\left[a_t + p_t\nabla_t  + \chi_t\right]
\end{equation*}
where $|\chi_t|\le \gamma + \frac{1}{2p(y_t| \alpha_t)|_{a_t}}\sup_{|\alpha_t-a_t|<\tilde{\delta}}\left |\frac{\partial^2 p(y_t|\alpha_t)}{\partial\alpha_t^2}\right |\left(\beta_t^{(3)}+|a_t|p_t\right)$. \\
Observe now that, since Assumption \eqref{ass:3} implies Assumption \eqref{ass:2}, Proposition \eqref{prop:pyt} holds and we can write:
\begin{align*}
a_{t|t}&= \frac{p(y_t| \alpha_t)|_{a_t}}{p(y_t| \alpha_t)|_{a_t} + \xi_t}\left(a_t + p_t\nabla_t  + \chi_t\right)\\
&= a_t + p_t\nabla_t  + \chi_t + \left(\frac{p(y_t| \alpha_t)|_{a_t}}{p(y_t| \alpha_t)|_{a_t} + \xi_t}-1\right)\left(a_t + p_t\nabla_t  + \chi_t\right)\\
&=  a_t + p_t\nabla_t  + \chi_t + \left(-\frac{\xi_t}{p(y_t| \alpha_t)|_{a_t} + \xi_t}\right)\left(a_t + p_t\nabla_t  + \chi_t\right)\\
&= a_t + p_t\nabla_t  + \chi_t + O(\xi_t) .
\end{align*}
 \begin{flushright}
$\square$
\end{flushright}

\section{Proof of Theorem \eqref{prop:ptt}}
\label{app:prop:ptt}
First, observe that:
\begin{equation}
p_{t|t} = \mathbb{E}(\alpha_t-a_{t|t})^2|Y_t] = \mathbb{E}(\alpha_t-a_t)^2|Y_t] - \left(a_{t|t}-a_t\right)^2
\label{eq:app:mtt}
\end{equation}
The first term can be written as:
\begin{equation*}
\mathbb{E}(\alpha_t-a_t)^2|Y_t] = \int_{-\infty}^{+\infty}(\alpha_t-a_t)^2\frac{p(y_t|\alpha_t)p(\alpha_t|Y_{t-1})}{p(y_t|Y_{t-1})}d\alpha_t
\end{equation*}
To compute this integral, we expand $p(y_t|\alpha_t)$ at second order in a neighborhood of center $a_t$ and radius $\delta$:
\begin{equation*}
p(y_t|\alpha_t) = p(y_t|\alpha_t)|_{a_t} + \frac{\partial p(y_t|\alpha_t)}{\partial\alpha_t}\bigg |_{a_t}\left(\alpha_t-a_t\right)+ \frac{1}{2}\frac{\partial^2 p(y_t|\alpha_t)}{\partial\alpha_t^2}\bigg |_{a_t}\left(\alpha_t-a_t\right)^2+  g(\alpha_t)
\end{equation*} 
where $|g(\alpha_t)|\le M(\delta)|\alpha_t-a_t|^3$ for each $\alpha_t$ belonging to the neighborhood. Thanks to Assumption \eqref{ass:4}, we have $M(\delta)=\frac{1}{6}\sup_{|\alpha_t-a_t|\le \delta} \left| \frac{\partial^3 p(y_t|\alpha_t)}{\partial\alpha_t^3}\right|$. Moreover, we set $\delta\ge \bar{\delta}$, where $\bar{\delta}$ is defined in Assumption \eqref{ass:5}. We can thus write:
\begin{align*}
\mathbb{E}(\alpha_t-a_t)^2|Y_t] &= \frac{p(y_t|\alpha_t)|_{a_t}}{p(y_t|Y_{t-1})}\int_{-\infty}^{+\infty}\left(\alpha_t-a_t\right)^2\left[1+\nabla_t\left(\alpha_t-a_t\right) + \right. \\
& \left. +
\frac{1}{2p(y_t|\alpha_t)|_{a_t}}\frac{\partial^2 p(y_t|\alpha_t)}{\partial\alpha_t^2}\bigg |_{a_t}\left(\alpha_t-a_t\right)^2+ \frac{g(\alpha_t)}{p(y_t|\alpha_t)|_{a_t}}\right]    p(\alpha_t|Y_{t-1})d\alpha_t
\end{align*}
The integral of the first three terms results in:
\begin{align}
&\quad \int_{-\infty}^{+\infty}(\alpha_t-a_t)^2\left[1+\nabla_t\left(\alpha_t-a_t\right) +
\frac{1}{2p(y_t|\alpha_t)|_{a_t}}\frac{\partial^2 p(y_t|\alpha_t)}{\partial\alpha_t^2}\bigg |_{a_t}\left(\alpha_t-a_t\right)^2\right]    p(\alpha_t|Y_{t-1})d\alpha_t \nonumber \\
& = p_t + \nabla_t p_t^{(3)}+ \frac{1}{2p(y_t|\alpha_t)|_{a_t}}\frac{\partial^2 p(y_t|\alpha_t)}{\partial\alpha_t^2}\bigg |_{a_t}p_t^{(4)}\\
&=p_t + l_t
\label{eq:app:ppt1}
\end{align}
where $p_t^{(3)}$ and $p_t^{(4)}$ denote the third and fourth moments of $ p(\alpha_t|Y_{t-1})$, respectively, and $l_t=\nabla_t p_t^{(3)}+ \frac{1}{2p(y_t|\alpha_t)|_{a_t}}\frac{\partial^2 p(y_t|\alpha_t)}{\partial\alpha_t^2}\bigg |_{a_t}p_t^{(4)}$. 

We now compute the second term in Equation \eqref{eq:app:mtt}. Observe that Proposition \eqref{prop:pyt} and Theorem \eqref{prop:att} hold under Assumptions \eqref{ass:4}, \eqref{ass:5}. Thus we can write:
\begin{align*}
\left(a_{t|t}-a_t\right)^2&=\left[\frac{p(y_t| \alpha_t)|_{a_t}}{p(y_t|Y_{t-1})}\left(a_t + p_t\nabla_t  + \chi_t\right)-a_t\right]^2\\
&= \left[\left(\frac{p(y_t|\alpha_t)|_{a_t}}{p(y_t|Y_{t-1})}-1\right)a_t + \frac{p(y_t|\alpha_t)|_{a_t}}{p(y_t|Y_{t-1})}p_t\nabla_t+ \frac{p(y_t|\alpha_t)|_{a_t}}{p(y_t|Y_{t-1})}\chi_t  \right]^2\\
&=\left[-\frac{\xi_t}{p(y_t|\alpha_t)|_{a_t}+\xi_t}a_t  + \frac{p(y_t|\alpha_t)|_{a_t}}{p(y_t|Y_{t-1})}p_t\nabla_t+ \frac{p(y_t|\alpha_t)|_{a_t}}{p(y_t|Y_{t-1})}\chi_t \right]^2
\end{align*}
The leading term in the above equation is the second one, whereas the first and the third terms are of order $\xi_t$ and $\chi_t$, respectively.
The square of the leading term can be written as:
\begin{align*}
&\quad \left[\frac{p(y_t|\alpha_t)|_{a_t}}{p(y_t|Y_{t-1})}\right]^2\left(p_t\right)^2\left(\nabla_t\right)^2 \\
& = \frac{p(y_t|\alpha_t)|_{a_t}}{p(y_t|Y_{t-1})}\left(p_t\right)^2\left(\nabla_t\right)^2\left[\frac{p(y_t|\alpha_t)|_{a_t}}{p(y_t|Y_{t-1})}-1\right]+ \frac{p(y_t|\alpha_t)|_{a_t}}{p(y_t|Y_{t-1})}\left(p_t\right)^2\left(\nabla_t\right)^2\\
& = - \frac{p(y_t|\alpha_t)|_{a_t}}{p(y_t|Y_{t-1})}\left(p_t\right)^2\left(\nabla_t\right)^2\frac{\xi_t}{p(y_t|Y_{t-1})}+ \frac{p(y_t|\alpha_t)|_{a_t}}{p(y_t|Y_{t-1})}\left(p_t\right)^2\left(\nabla_t\right)^2
\end{align*}
Combining this result with Equation \eqref{eq:app:ppt1}, we can write $p_{t|t}$ as:
\begin{align*}
p_{t|t} &=  \frac{p(y_t|\alpha_t)|_{a_t}}{p(y_t|Y_{t-1})}\left[p_t  - \left(p_t\right)^2\left(\nabla_t\right)^2+ \ell_t\right] \\
& + O(\xi_t) + O(\chi_t) +  \frac{p(y_t|\alpha_t)|_{a_t}}{p(y_t|Y_{t-1})}\int_{-\infty}^{+\infty}(\alpha_t-a_t)^2\frac{g(\alpha_t)}{{p(y_t|\alpha_t)|_{a_t}}}p(\alpha_t|Y_{t-1})d\alpha_t
\end{align*}
where $O(\xi_t)$, $O(\chi_t)$ denote higher order terms in $\xi_t$ and $\chi_t$. 
To conclude the proof, we need to bound the integral in $g(\alpha_t)$. Let us decompose this integral as\footnote{The term $\frac{1}{p(y_t|\alpha_t)_{a_t}}$ will be absorbed in the definition of $\tilde{\delta}$ below.}:
\begin{align*}
&\quad \int_{-\infty}^{+\infty}(\alpha_t-a_t)^2g(\alpha_t)p(\alpha_t|Y_{t-1})d\alpha_t = \int_{-\infty}^{a_t-\delta}(\alpha_t-a_t)^2g(\alpha_t)p(\alpha_t|Y_{t-1})d\alpha_t\\
&+ \int_{a_t-\delta}^{a_t+\delta}(\alpha_t-a_t)^2g(\alpha_t)p(\alpha_t|Y_{t-1})d\alpha_t+ \int_{a_t+\delta}^{+\infty}(\alpha_t-a_t)^2g(\alpha_t)p(\alpha_t|Y_{t-1})d\alpha_t
\end{align*}
The integral over the neighborhood of center $a_t$ and radius $\delta$ can be bounded thanks to Assumption \eqref{ass:4}:
\begin{align*}
\left|\int_{a_t-\delta}^{a_t+\delta}(\alpha_t-a_t)^2g(\alpha_t)p(\alpha_t|Y_{t-1})d\alpha_t\right| &\le 
\int_{a_t-\delta}^{a_t+\delta}(\alpha_t-a_t)^2|g(\alpha_t)|p(\alpha_t|Y_{t-1})d\alpha_t\\
&= \int_{a_t-\delta}^{a_t+\delta} M(\delta) |\alpha_t-a_t|^5p(\alpha_t|Y_{t-1})d\alpha_t\\
&\le M(\delta) \beta_t^{(5)}
\end{align*}
The two integrals outside the neighborhood can instead be bounded using Assumption \eqref{ass:5} and the boundedness of $p(y_t|\alpha_t)$ in Assumption \eqref{ass:1}. Let us focus first on the third integral:
\begin{align*}
&\quad \left| \int_{a_t+\delta}^{+\infty}(\alpha_t-a_t)^2g(\alpha_t)p(\alpha_t|Y_{t-1})d\alpha_t\right|\\
&\le \int_{a_t+\delta}^{+\infty}(\alpha_t-a_t)^2|g(\alpha_t)|p(\alpha_t|Y_{t-1})d\alpha_t\\
&= \int_{a_t+\delta}^{+\infty}(\alpha_t-a_t)^2\left| p(y_t|\alpha_t)-p(y_t|\alpha_t)|_{a_t} - \frac{\partial p(y_t|\alpha_t)}{\partial\alpha_t}\bigg |_{a_t}\left(\alpha_t-a_t\right) \right.\\
& \left. - \frac{1}{2}\frac{\partial^2 p(y_t|\alpha_t)}{\partial\alpha_t^2}\bigg |_{a_t}\left(\alpha_t-a_t\right)^2   \right|p(\alpha_t|Y_{t-1})d\alpha_t\\
&\le \int_{a_t+\delta}^{+\infty}\frac{\sup_{\alpha_t>a_t+\delta}p(y_t|\alpha_t)+p(y_t|\alpha_t)|_{a_t}}{(\alpha_t-a_t)^{4+\beta}}d\alpha_t+ \int_{a_t+\delta}^{+\infty}\frac{\left|\frac{\partial p(y_t|\alpha_t)}{\partial\alpha_t}\bigg |_{a_t}\right|}{(\alpha_t-a_t)^{3+\beta}}d\alpha_t\\
&+  \int_{a_t+\delta}^{+\infty}\frac{\left|\frac{\frac{1}{2}\partial^2 p(y_t|\alpha_t)}{\partial\alpha_t^2}\bigg |_{a_t}\right|}{(\alpha_t-a_t)^{2+\beta}}d\alpha_t\\
&= \frac{1}{3+\beta}\frac{\sup_{\alpha_t>a_t+\delta}p(y_t|\alpha_t)+p(y_t|\alpha_t)|_{a_t}}{\delta^{3+\beta}}+\frac{1}{2+\beta}\left|\frac{\partial p(y_t|\alpha_t)}{\partial\alpha_t}\bigg |_{a_t}\right|\frac{1}{\delta^{2+\beta}}\\
& + \frac{1}{1+\beta}\left|\frac{\frac{1}{2}\partial^2 p(y_t|\alpha_t)}{\partial\alpha_t^2}\bigg |_{a_t}\right|\frac{1}{\delta^{1+\beta}}
\end{align*}
The integral in $[-\infty,a_t]$ can be bounded in a similar way. The proof proceeds as in Section \eqref{app:prop:att}, i.e., by choosing $\tilde{\delta}$ as the lowest $\delta\ge \bar{\delta}$ for which the sum of the two integrals outside the neighborhood of center $a_t$ is in absolute value lower that $\gamma$. Setting $\zeta_t=\int_{-\infty}^{+\infty}(\alpha_t-a_t)^2\frac{g(\alpha_t)p(\alpha_t|Y_{t-1})}{p(y_t|\alpha_t)_{a_t}}d\alpha_t$, we thus have:
\begin{align*}
p_{t|t} =  \frac{p(y_t|\alpha_t)|_{a_t}}{p(y_t|Y_{t-1})}\left[p_t - \left(p_t\right)^2\nabla_t^2+ \ell_t + \zeta_t\right] + O(\xi_t) + O(\chi_t) 
\end{align*}
where $|\zeta_t|\le \gamma + \frac{1}{6p(y_t|\alpha_t)_{a_t}}\sup_{|\alpha_t-a_t|<\delta} \left| \frac{\partial^3 p(y_t|\alpha_t)}{\partial\alpha_t^3}\right|\beta_t^{(5)}$.
To conclude the proof, observe that if Assumptions $\eqref{ass:4}$ and $\eqref{ass:5}$ are satisfied, than Proposition \eqref{prop:pyt} holds and we can write:
\begin{align*}
p_{t|t} &= \frac{p(y_t|\alpha_t)|_{a_t}}{p(y_t|\alpha_t)|_{a_t}+\xi_t}\left[p_t - \left(p_t\right)^2\nabla_t^2+ \ell_t + \zeta_t\right] + O(\xi_t) + O(\chi_t) \\
&= p_t - \left(p_t\right)^2\nabla_t^2+ \ell_t + \zeta_t -\frac{\xi_t}{p(y_t|\alpha_t)|_{a_t}+\xi_t}[p_t + \left(p_t\right)^2h_t+ \ell_t + \zeta_t] + O(\xi_t) + O(\chi_t) \\
& =  p_t - \left(p_t\right)^2\nabla_t^2+ \ell_t + \zeta_t + O(\xi_t) + O(\chi_t).
\end{align*}
 \begin{flushright}
$\square$
\end{flushright}

\section{Proof of Proposition \eqref{prop:smooth}}
\label{app:prop:smoothG}

This result is a consequence of the two-filter formula for smoothing; see \cite{Genshiro94}. Observe that we can write, for $t\le n$:
\begin{align*}
p(\alpha_t|Y_n)=p(\alpha_t|y_1,\dots,y_n)&=
\frac{p(\alpha_t,y_t,\dots,y_n|y_1,\dots,y_{t-1})}{p(y_t,\dots,y_n|y_1,\dots,y_{t-1})}\\
&=\frac{p(y_t,\dots,y_n|\alpha_t)p(\alpha_t|y_1,\dots,y_{t-1})}{p(y_t,\dots,y_n|y_1,\dots,y_{t-1})}\\
&=\frac{p(y_t,\dots,y_n|\alpha_t)p(\alpha_t|Y_{t-1})}{p(y_t,\dots,y_n|y_1,\dots,y_{t-1})}
\end{align*}
The term $p(y_t,\dots,y_n|\alpha_t)$ can be decomposed as follows:
\begin{align*}
p(y_t,\dots,y_n|\alpha_t) &= p(y_{t+1},\dots,y_n|\alpha_t)p(y_t|\alpha_t)\\
&= \int_{-\infty}^{+\infty}p(y_{t+1},\dots,y_n,\alpha_{t+1}|\alpha_t)p(y_t|\alpha_t)d\alpha_{t+1}\\
&= \int_{-\infty}^{+\infty}p(y_{t+1},\dots,y_n|\alpha_{t+1})p(\alpha_{t+1}|\alpha_t)p(y_t|\alpha_t)d\alpha_{t+1}
\end{align*}
Using the same decomposition for $p(y_{t+1},\dots,y_n|\alpha_{t+1})$ and iterating several times, we obtain:
\begin{align*}
p(y_t,\dots,y_n|\alpha_t) = \int_{-\infty}^{+\infty}\dots \int_{-\infty}^{+\infty}\prod_{i=t}^n p(y_i|\alpha_i)\prod_{j=t+1}^n p(\alpha_j|\alpha_{j-1})d\alpha_{t+1}\dots d\alpha_n
\end{align*}
The $\ell$-th moment of $p(\alpha_t|Y_n)$ is then:
\begin{align*}
\mathbb{E}[\alpha_t^{\ell}|Y_n] &= \int_{-\infty}^{+\infty} \alpha_t^{\ell}p(\alpha_t|Y_n)d\alpha_t\\
& = N_t'\int_{-\infty}^{+\infty}\dots \int_{-\infty}^{+\infty}\alpha_t^{\ell}\left(\prod_{i=t}^n p(y_i|\alpha_i)\prod_{j=t+1}^n p(\alpha_j|\alpha_{j-1})\right)p(\alpha_t|Y_{t-1}) d\alpha_{t}\dots d\alpha_n
\end{align*}
where $N_t'=p(y_t,\dots,y_n|y_1,\dots,y_{t-1})^{-1}$. The latter expressione can be written more compactly in terms of the sequence $\{K_i(\alpha_i)\}_{i=t}^{n-1}$ defined recursively as follows:
\begin{equation*}
 K_i(\alpha_i)=\int_{-\infty}^{+\infty} K_{i+1}(\alpha_{i+1})p(y_{i+1}|\alpha_{i+1})p(\alpha_{i+1}|\alpha_i)d\alpha_{i+1}
\end{equation*} 
for $i=t,t+1,\dots,n-2$, and $K_{n-1}(\alpha_{n-1})=\int_{-\infty}^{+\infty} p(y_{n}|\alpha_{n})p(\alpha_{n}|\alpha_{n-1})d\alpha_{n}$. In terms of $\{K_i(\alpha_i)\}_{i=t}^{n-1}$, we can write the smoothed moments as:
\begin{equation*}
\mathbb{E}[\alpha_t^{\ell}|Y_n] = \int_{-\infty}^{+\infty}\alpha_t^{\ell}K_t(\alpha_t)p(y_t|\alpha_t)p(\alpha_t|Y_{t-1})d\alpha_t
\end{equation*}

\begin{flushright}
$\square$
\end{flushright}

\end{document}